\documentclass[12pt,eqno,epsf]{article}
\usepackage{amsmath,amssymb,graphicx,slashbox}
\numberwithin{equation}{section}

\def\mydate{March 12, 2013}
\def\ignore#1{{}}

\newcounter{sxn}

\newcounter{axn}

\date{}

\newdimen\mybaselineskip
\renewcommand{\thefootnote}{\arabic{footnote}}

\newcommand{\beeq}{\begin{equation}}
\newcommand{\eneq}{\end{equation}}
\newcommand{\beqn}{\begin{eqnarray}}
\newcommand{\eeqn}{\end{eqnarray}}

\newcommand{\alp}{\alpha}
\newcommand{\bt}{\beta}
\newcommand{\gm}{\gamma}

\newcommand{\dlt}{\delta}
\newcommand{\Dlt}{\Delta}
\newcommand{\ep}{\epsilon}
\newcommand{\tht}{\theta}

\newcommand{\vth}{\vartheta}

\newcommand{\lmd}{\lambda}
\newcommand{\Lmd}{\Lambda}
\newcommand{\sgm}{\sigma}
\newcommand{\Sgm}{\Sigma}
\newcommand{\vph}{\varphi}
\newcommand{\omg}{\omega}
\newcommand{\Omg}{\Omega}
\newcommand{\dalp}{\dot{\alpha}}
\newcommand{\dbt}{\dot{\beta}}
\newcommand{\dgm}{\dot{\gamma}}

\newcommand{\be}{\begin{equation}}
\newcommand{\ee}{\end{equation}}
\newcommand{\bea}{\begin{eqnarray}}
\newcommand{\eea}{\end{eqnarray}}
\newcommand{\eql}{\!\!\!&=\!\!\!&}

\newcommand{\defa}{\!\!\!&\equiv\!\!\!&}

\newcommand{\toa}{\!\!\!&\to\!\!\!&}

\newcommand{\tl}[1]{\tilde{#1}}
\newcommand{\bdm}[1]{{\mbox{\boldmath $#1$}}}
\newcommand{\tr}{{\rm tr}}

\newcommand{\diag}{{\rm diag}}
\newcommand{\der}{\partial}
\newcommand{\dr}{\!\!d}
\newcommand{\hc}{{\rm h.c.}}
\newcommand{\ie}{{\it i.e.}}
\newcommand{\id}{\mbox{\boldmath $1$}}

\newcommand{\vev}[1]{\langle #1 \rangle}

\newcommand{\brkt}[1]{\left( #1 \right)}
\newcommand{\brc}[1]{\left\{ #1 \right\}}
\newcommand{\sbk}[1]{\left[ #1 \right]}
\newcommand{\abs}[1]{\left| #1 \right|}

\renewcommand{\Re}{{\rm Re}\,}
\renewcommand{\Im}{{\rm Im}\,}

\newcommand{\cA}{{\cal A}}

\newcommand{\cC}{{\cal C}}
\newcommand{\cD}{{\cal D}}
\newcommand{\cE}{{\cal E}}
\newcommand{\cF}{{\cal F}}
\newcommand{\cH}{{\cal H}}

\newcommand{\cL}{{\cal L}}

\newcommand{\cN}{{\cal N}}
\newcommand{\cO}{{\cal O}}

\newcommand{\cV}{{\cal V}}
\newcommand{\cU}{{\cal U}}
\newcommand{\cW}{{\cal W}}
\newcommand{\cX}{{\cal X}}
\newcommand{\cY}{{\cal Y}}
\newcommand{\cZ}{{\cal Z}} 
\newcommand{\bE}{{\mathbb E}}

\newcommand{\bH}{{\mathbb H}}

\newcommand{\dsc}{\delta_{\rm sc}}
\newcommand{\dscp}{\delta_{\rm sc}^{(1)}}
\newcommand{\dscq}{\delta_{\rm sc}^{(2)}}
\newcommand{\dgt}{\delta_{\rm gauge}}
\newcommand{\suU}{SU(2)_U}
\newcommand{\oa}{\overline{a}}
\newcommand{\ob}{\overline{b}}
\newcommand{\oc}{\overline{c}}
\newcommand{\od}{\overline{d}}

\begin{document}
\thispagestyle{empty}

\baselineskip=12pt

{\small \noindent \mydate    
\hfill }

{\small \noindent \hfill  KEK-TH-1544}

\baselineskip=35pt plus 1pt minus 1pt

\vskip 1.5cm

\begin{center}
{\LARGE\bf Superfield description of gravitational} \\
{\LARGE\bf couplings in generic 5D supergravity}

\vspace{1.5cm}
\baselineskip=20pt plus 1pt minus 1pt

\normalsize

{\large\bf Yutaka\ Sakamura}$\!${\def\thefootnote{\fnsymbol{footnote}}
\footnote[1]{\tt e-mail address: sakamura@post.kek.jp}}

\vspace{.3cm}
{\small \it KEK Theory Center, Institute of Particle and Nuclear Studies, 
KEK, \\ Tsukuba, Ibaraki 305-0801, Japan} \\ \vspace{3mm}
{\small \it Department of Particles and Nuclear Physics, \\
The Graduate University for Advanced Studies (Sokendai), \\
Tsukuba, Ibaraki 305-0801, Japan} 
\end{center}

\vskip 1.0cm
\baselineskip=20pt plus 1pt minus 1pt

\begin{abstract}
We complete the $\cN=1$ superfield action for the generic system of 
vector multiplets and hypermultiplets coupled to 5D supergravity, 
which is based on the superconformal formulation. 
Especially we clarify the gravitational couplings to the bulk matters 
at linear order in the gravitational superfields. 
They consist of four $\cN=1$ superfields, 
two of which are $Z_2$-odd when the fifth dimension 
is compactified on $S^1/Z_2$. 
This formulation provides a powerful tool to calculate quantum effects, 
keeping the $\cN=1$ off-shell structure. 
\end{abstract}


\newpage

\section{Introduction}
Higher dimensional supergravity (SUGRA) has been attracted much attention 
and extensively investigated in various aspects, such as effective theories 
of the superstring theory or M-theory, 
AdS/CFT correspondence~\cite{Maldacena:1997re}, 
the model building in the context of the brane-world scenario, etc. 
Among them, five-dimensional (5D) SUGRA 
compactified on an orbifold~$S^1/Z_2$ has been thoroughly investigated 
since it is shown to appear as an effective theory of the strongly coupled 
heterotic string theory~\cite{Horava:1995qa} 
compactified on a Calabi-Yau 3-fold~\cite{Lukas:1998yy}. 
Besides, the supersymmetric (SUSY) extensions of 
the Randall-Sundrum model~\cite{Randall:1999ee}
are also constructed in 5D SUGRA 
on $S^1/Z_2$~\cite{Gherghetta:2000qt,Falkowski:2000er,Altendorfer:2000rr}. 

The general form of 5D SUGRA in the component field formalism 
has been clarified in the following papers. 
The couplings of vector multiplets, tensor multiplets 
and hypermultiplets to SUGRA are discussed 
in Refs.~\cite{Gunaydin:1983bi}, \cite{Gunaydin:1999zx} 
and \cite{Ceresole:2000jd}, respectively. 
The results obtained in these works are reproduced 
in the superconformal formulation~\cite{Bergshoeff:2002qk}-\cite{Kugo:2002vc}, 
which is a systematic method to construct generic off-shell action 
of 5D SUGRA. 

Superfield formulation is an elegant and systematic way 
to construct SUSY theories. 
For five and six dimensions, 
the projective superspace approach offers nice off-shell 
formulations of SUGRA~\cite{Kuzenko:2007cj,Linch:2012zh}. 
On the other hand, such full superfield formulation 
is generically complicated and requires many auxiliary fields 
because of $\cN=2$ SUSY.\footnote{
$\cN=1$ denotes SUSY with four supercharges in this paper. }

It is useful and convenient to express higher dimensional SUSY theories 
in terms of $\cN=1$ superfields. 
Such superfield formulation enables us 
to describe interactions between fields localized on the brane 
and those in the bulk in a transparent manner. 
It also makes it easier to derive the low-energy four-dimensional (4D) 
effective theory that preserves $\cN=1$ SUSY. 
In the global SUSY case, such description of 
5-10 dimensional SUSY theories were provided in Ref.~\cite{ArkaniHamed:2001tb}. 
Especially 5D SUSY theories have been extensively 
investigated by the $\cN=1$ superfield 
description~\cite{Marti:2001iw,Hebecker:2001ke,Kuzenko:2005sz}. 
There are also some works along this direction for 5D SUGRA. 
In Ref.~\cite{Linch:2002wg}, the linearized minimal 5D SUGRA 
is constructed in $\cN=1$ superspace. 
This formulation is useful to calculate quantum loop effects 
from the gravitational fields propagating in the 5D bulk. 
However it is unclear how to extend their formulation to more generic case 
in which matter fields also propagate in the bulk.\footnote{
We refer to fields other than the gravitational fields 
as matters in this paper. } 

Another $\cN=1$ description of 5D SUGRA is based on 
the superconformal formulation. 
Since each 5D superconformal multiplet can be decomposed into 
$\cN=1$ multiplets~\cite{Kugo:2002js}, 
it is possible to express the generic 5D SUGRA action 
in terms of $\cN=1$ superfields. 
This has been done in Ref.~\cite{Abe:2004ar,Paccetti:2004ri}, 
and used to derive the 4D effective 
theories~\cite{Abe:2005ac}-\cite{Abe:2011rg}. 
However, the superfield action in these works is not a complete one. 
They did not take into account the $Z_2$-odd part of 
the gravitatioinal multiplet, and some terms involving the $Z_2$-even 
gravitational fields are also missing. 
These deficits are irrelevant if we focus on low-energy observables calculated  
at the classical level 
since $Z_2$-odd fields do not have zero-modes that appear 
in low-energy effective theory. 
However, when we discuss quantum loop effects, 
we have to take into account all fields in the theory, 
including the $Z_2$-odd fields. 
The purpose of this paper is to extend the results 
of Ref.~\cite{Abe:2004ar,Paccetti:2004ri} to include all the gravitational fields, 
and complete the $\cN=1$ superfield action for the generic system 
of vector multiplets and hypermultiplets coupled to 5D SUGRA.\footnote{
In this paper, we will not consider 5D tensor multiplets, 
which are discussed in Ref.~\cite{Gunaydin:1999zx,Kugo:2002vc}, 
just for simplicity. 
For practical purposes, our setup is useful enough  
because we can construct various phenomenological models without them. 
It is possible to incorporate the 5D tensor multiplets 
into our formulation by using the results in Ref.~\cite{Kugo:2002vc}. 
} 
In this paper, we keep terms up to linear order 
in the gravitational superfields for each interaction term. 
This is enough for one-loop calculations. 
Thus our work can also be understood as an extension of Ref.~\cite{Linch:2002wg} 
to a case that the matter superfields propagate in the bulk. 

The paper is organized as follows. 
In the next two sections, we review our previous works. 
We provide the superfield description of 4D SUGRA 
based on the superconformal formulation in Sec.~\ref{4D_SUGRA}, and  
that of 5D SUGRA in which the gravitational fluctuation fields are dropped
in Sec.~\ref{5D_SUGRA}. 
In Sec.~\ref{grav_cp}, the gravitational superfields are introduced 
as the connections for the 5D superconformal symmetries. 
Their couplings to the matter superfields are determined 
from the invariance of the action. 
Sec.~\ref{summary} is devoted to the summary. 
In Appendix~\ref{Trf:fss}, we check that 4D field strength superfields 
defined in Sec.~\ref{4D_SUGRA} 
correctly transform as superconformal chiral multiplets. 
In Appendix~\ref{comp:superfields}, we collect explicit expressions 
of $\cN=1$ matter superfields in terms of component fields 
of 5D superconformal multiplets. 
In Appendix~\ref{action_inv}, we show the invariance of the action 
under the supergauge and the $Z_2$-odd superconformal transformations 
in the superfield description.

\section{4D supergravity} \label{4D_SUGRA} 
In this section, we review our previous work~\cite{Sakamura:2011df} 
that derives the superfield description of 4D SUGRA 
based on the superconformal formulation of Ref.~\cite{Kugo:1982cu}. 
This is a modified version of the superfield formalism 
in Ref.~\cite{Siegel:1978mj} that makes  
a relation to the formalism of Ref.~\cite{Kugo:1982cu} manifest. 
We will use formulae in this section 
to express 5D SUGRA action in Sec.~\ref{grav_cp}. 

We assume that the background geometry is a flat 4D Minkowski spacetime. 
Basically we use the two-component spinor notations of Ref.~\cite{Wess:1992cp}, 
except for the metric and the spinor derivatives. 
We take the background metric as $\eta_{\mu\nu}=\diag(1,-1,-1,-1)$ 
so as to match it to that of Ref.~\cite{Kugo:1982cu}, and we define the spinor 
derivatives~$D_\alp$ and $\bar{D}_{\dalp}$ as 
\be
 D_\alp \equiv \frac{\der}{\der\tht^\alp}
 -i\brkt{\sgm^\mu\bar{\tht}}_\alp\der_\mu, \;\;\;\;\;
 \bar{D}_{\dalp} \equiv -\frac{\der}{\der\bar{\tht}^{\dalp}}
 +i\brkt{\tht\sgm^\mu}_{\dalp}\der_\mu, 
\ee
which satisfy $\brc{D_\alp,\bar{D}_{\dalp}}=2i\sgm^\mu_{\alp\dalp}\der_\mu$.

\subsection{Definition of superfields} \label{def:sf:4D}
The 4D superconformal algebra consists of 
the translation~$\bdm{P}$, the Lorentz transformation~$\bdm{M}$, 
SUSY~$\bdm{Q}$, the R symmetry~$U(1)_A$, the dilatation~$\bdm{D}$, 
the conformal SUSY~$\bdm{S}$ and the conformal boost~$\bdm{K}$. 
Among the gauge fields for these symmetries, 
only the vierbein~$e_\mu^{\;\;\underline{\nu}}$, 
the gravitino~$\psi_\mu$, the $U(1)_A$ gauge field~$A_\mu$ 
and the $\bdm{D}$ gauge field~$b_\mu$ are independent 
degrees of freedom~\cite{Kugo:1982cu}. 
In our previous work~\cite{Sakamura:2011df}, 
we showed that these fields form the following real superfield 
with an external Lorentz index. 
\be
 U^\mu = \brkt{\tht\sgm^\nu\bar{\tht}}\tl{e}_\nu^{\;\;\mu}
 +i\bar{\tht}^2\brkt{\tht\sgm^\nu\bar{\sgm}^\mu\psi_\nu}
 -i\tht^2\brkt{\bar{\tht}\bar{\sgm}^\nu\sgm^\mu\bar{\psi}_\nu}
 +\frac{1}{4}\tht^2\bar{\tht}^2\brkt{3A^\mu-\ep^{\mu\nu\rho\tau}
 \der_\nu\tl{e}_{\rho\tau}},  
\ee
where $\tl{e}_\nu^{\;\;\mu}\equiv e_\nu^{\;\;\underline{\mu}}-\dlt_\nu^{\;\;\mu}$ 
is the fluctuation around the background.\footnote{
In the formulation of Ref.~\cite{Kugo:1982cu}, 
the $\bdm{D}$ gauge field~$b_\mu$ does not play any essential role, 
and can be set to zero. 
This corresponds to the gauge fixing condition for $\bdm{K}$. \label{b_zero}}  
In our formulation, we keep the gravitational 
fields~$(\tl{e}_\mu^{\;\;\nu},\psi_\mu,A_\mu)$ up to linear order. 
Therefore we need not discriminate the curved indices~$\mu,\nu,\cdots$ 
from the flat ones~$\underline{\mu},\underline{\nu},\cdots$ 
since the background geometry is flat. 
Thus we omit the underlines of the flat indices in the following. 

A (superconformal) chiral multiplet~$\sbk{\phi,\chi_\alp,F}$ 
is expressed by the following chiral superfield. 
\be
 \Phi = \brkt{1+\frac{w}{3}\cE}\brkt{\phi+\tht\chi+\tht^2F}, 
 \label{sf:chiral}
\ee
where $w$ is the Weyl weight (\ie, the $\bdm{D}$ charge) of this multiplet,\footnote{
The Weyl weight of a multiplet denotes that of the lowest component 
in the multiplet. } 
and 
\be
 \cE \equiv \tl{e}_\mu^{\;\;\mu}-2i\tht\sgm^\mu\bar{\psi}_\mu, \label{def:cE}
\ee
corresponds to the fluctuation part of the chiral density multiplet 
in Ref.~\cite{Wess:1992cp}.  
We have worked in the chiral 
coordinate~$y^\mu\equiv x^\mu-i\tht\sgm^\mu\bar{\tht}$ 
to express these chiral superfields. 
In the superconformal formulation of Ref.~\cite{Kugo:1982cu}, 
there is a formula that embeds a chiral multiplet into a general multiplet. 
It is expressed in our superfield description as 
\be
 \cU(\Phi) \equiv \brkt{1+iU^\mu\der_\mu}\Phi, \;\;\;\;\;
 \cU(\bar{\Phi}) \equiv \brkt{1-iU^\mu\der_\mu}\bar{\Phi}.  \label{def:cU}
\ee
Each chiral superfield~$\Phi$ in the full superspace integral~$\int\dr^4\tht$ 
must appear in this form. 

A real general multiplet~$\sbk{C,\zeta_\alp,\cH,B_\mu,\lmd_\alp,D}$
is expressed~\footnote{
A complex scalar~$\cH$ should be understood as $\frac{1}{2}\brkt{H+iK}$ 
in the notation of Ref.~\cite{Kugo:1982cu}. } 
by a real superfields, 
\bea
 V \eql \brc{1+\frac{w}{6}\brkt{\cE+\bar{\cE}}} 
 \left\{C+i\tht\zeta-i\bar{\tht}\bar{\zeta}-\tht^2\cH-\bar{\tht}^2\bar{\cH}
 -\brkt{\tht\sgm^\mu\bar{\tht}}B'_\mu \right. \nonumber\\
 &&\left. \hspace{35mm}
 +i\tht^2\brkt{\bar{\tht}\bar{\lmd}'}
 -i\bar{\tht}^2\brkt{\tht\lmd'}
 +\frac{1}{2}\tht^2\bar{\tht}^2D' \right\}, 
\eea
where $w$ is the Weyl weight of this multiplet, and 
\bea
 B'_\mu \defa B_\mu-\zeta\psi_\mu-\bar{\zeta}\bar{\psi}_\mu
 -\frac{w}{2}C A_\mu, \nonumber\\
 \lmd'_\alp \defa \lmd_\alp-\frac{i}{2}\brc{\sgm^\mu\brkt{e^{-1}}_\mu^{\;\;\nu}
 \der_\nu\bar{\zeta}}_\alp-\brkt{\sgm^\mu\bar{\sgm}^\nu\psi_\mu}_\alp B_\nu
 -\frac{w}{4}\brkt{\sgm^\mu\bar{\zeta}}_\alp A_\mu, \nonumber\\
 D' \defa D-\frac{1}{2}g^{\mu\nu}\der_\mu\der_\nu C
 -\brkt{\bar{\lmd}\bar{\sgm}^\mu\psi_\mu
 -\frac{i}{2}\der_\nu\zeta\sgm^\mu\bar{\sgm}^\nu\psi_\mu
 -i\der_\mu\zeta\psi^\mu-\frac{2iw}{3}\zeta\sgm^{\mu\nu}\der_\nu\psi_\mu+\hc} 
 \nonumber\\
 &&+\brkt{\frac{3-w}{2}A^\mu-\frac{1}{2}\ep^{\mu\nu\rho\tau}
 \der_\nu\tl{e}_{\rho\tau}}B_\mu.  \label{comp:V}
\eea
Here $\brkt{e^{-1}}_\mu^{\;\;\nu}\equiv\dlt_\mu^{\;\;\nu}-\tl{e}_\mu^{\;\;\nu}$
and $g^{\mu\nu}\equiv\eta^{\mu\nu}-\tl{e}^{\mu\nu}-\tl{e}^{\nu\mu}$ are 
the inverse matrices of the vierbein and the metric, respectively. 

The gauge multiplet is a real multiplet with $w=0$. 
The gauge field~$\hat{B}_\mu$ is identified with 
\be
 \hat{B}_\mu \equiv \brkt{\dlt_\mu^{\;\;\nu}+\tl{e}_\mu^{\;\;\nu}}B'_\nu 
 = \brkt{\dlt_\mu^{\;\;\nu}+\tl{e}_\mu^{\;\;\nu}}B_\nu
 -\zeta\psi_\mu-\bar{\zeta}\bar{\psi}_\mu. 
\ee
For simplicity, we consider a case of the Abelian gauge group. 
An extension to the non-Abelian case is straightforward 
as explained in Sec.~\ref{fss}. 
The (super)gauge transformation is expressed in our superfield description as 
\be
 V \to V+\cU(\Lmd)+\cU(\bar{\Lmd}),  \label{ab:gauge_trf}
\ee
where the transformation parameter~$\Lmd=\phi^\Lmd+\tht\chi^\Lmd+\tht^2F^\Lmd$ 
(in the $y^\mu$-coordinate) is a chiral superfield. 
Note that $\Lmd$ must be embedded into a general multiplet by $\cU$ 
in order to be added to $V$. 
We can move to the Wess-Zumino gauge by choosing $\Lmd$ as~\footnote{
This gauge is possible only in the case of $w=0$. } 
\be
 \Re\phi^\Lmd = -\frac{1}{2}C, \;\;\;\;\;
 \chi^\Lmd_\alp = -i\zeta_\alp, \;\;\;\;\;
 F^\Lmd = \cH. 
\ee
In this gauge, $V$ is written as 
\bea
 V_{\rm WZ} \eql -\brkt{\tht\sgm^\mu\bar{\tht}}\brkt{e^{-1}}_\mu^{\;\;\nu}
 \hat{B}_\nu+i\tht^2\bar{\tht}\brc{\bar{\lmd}
 -\brkt{\bar{\sgm}^\nu\sgm^\mu\bar{\psi}_\nu}\hat{B}_\mu}
 -i\bar{\tht}^2\tht\brc{\lmd-\brkt{\sgm^\nu\bar{\sgm}^\mu\psi_\nu}\hat{B}_\mu}
 \nonumber\\
 &&+\frac{1}{2}\tht^2\bar{\tht}^2\brc{D
 -\brkt{\bar{\lmd}\bar{\sgm}^\mu\psi_\mu+\hc}
 +\brkt{\frac{3}{2}A^\mu-\frac{1}{2}\ep^{\mu\nu\rho\tau}\der_\nu\tl{e}_{\rho\tau}}
 \hat{B}_\mu}, 
\eea
where $\hat{B}_\mu$ is understood as the gauge-transformed gauge field. 
A set of the components$\sbk{\hat{B}_\mu,\lmd_\alp,D}$ 
form a gauge multiplet.

\subsection{Superconformal transformation}
Throughout this paper, we neglect terms beyond linear order 
in $U^\mu$ in the action, except for its kinetic terms 
that are discussed in Sec.~\ref{L_kin:U}. 
Hence it is enough to ensure an invariance of the action 
under the superconformal transformations up to the zeroth order in $U^\mu$ 
because it transforms inhomogeneously.  
The (linearized) superconformal transformations of the superfields 
defined in the previous subsection are expressed as 
\bea
 \dsc U^\mu \eql \frac{1}{2}\sgm^\mu_{\alp\dalp}
 \brkt{\bar{D}^{\dalp}L^\alp-D^\alp\bar{L}^{\dalp}}, \nonumber\\
 \dsc\Phi \eql \brc{-\frac{1}{4}\bar{D}^2L^\alp D_\alp
 -i\sgm^\mu_{\alp\dalp}\bar{D}^{\dalp}L^\alp\der_\mu
 -\frac{w}{12}\bar{D}^2D^\alp L_\alp}\Phi, \nonumber\\
 \dsc V \eql \brc{-\frac{1}{4}\bar{D}^2L^\alp D_\alp
 -\frac{i}{2}\sgm^\mu_{\alp\dalp}\bar{D}^{\dalp}L^\alp\der_\mu
 -\frac{w}{24}\bar{D}^2D^\alp L_\alp+\hc}V,  \label{dsc_trf}
\eea
where $L_\alp$ is a transformation parameter superfield.\footnote{
We have set $\Omg^\mu$ in Ref.~\cite{Sakamura:2011df} to zero. 
This is always possible by imposing constraints on the components of $L_\alp$. 
} 
Define 
\bea
 \xi^\mu \defa -\left.\Re\brkt{i\sgm^\mu_{\alp\dalp}\bar{D}^{\dalp}L^\alp}
 \right|_0, \;\;\;\;\;
 \ep_\alp \equiv -\frac{1}{4}\left.\bar{D}^2L_\alp\right|_0, \nonumber\\
 \lmd_{\mu\nu} \defa -\frac{1}{2}\left.\Re\brc{
 \brkt{\sgm_{\mu\nu}}_\bt^{\;\;\alp}D_\alp\bar{D}^2L^\bt}\right|_0, \;\;\;\;\;
 \vph_D \equiv \left.\Re\brkt{\frac{1}{4}D^\alp\bar{D}^2L_\alp}\right|_0, 
 \nonumber\\
 \vth_A \defa \left.\Im\brkt{-\frac{1}{6}D^\alp\bar{D}^2L_\alp}\right|_0, 
 \;\;\;\;\;
 \eta_\alp \equiv -\frac{1}{32}\left.D^2\bar{D}^2L_\alp\right|_0, 
\eea
where the symbol~$|_0$ denotes the lowest component of a superfield.  
Then these components are identified with the transformation parameters 
for $\bdm{P}$, $\bdm{Q}$, $\bdm{M}$, $\bdm{D}$, $U(1)_A$ and $\bdm{S}$, 
respectively. 
We have explicitly checked that (\ref{dsc_trf}) reproduces the correct 
superconformal transformations of each component field 
listed in Ref.~\cite{Kugo:1982cu}.

\subsection{Field strength superfield} \label{fss}
In the Abelian case, we define 
\be
 X \equiv \brkt{1+\frac{1}{4}U^\mu\bar{\sgm}^{\dalp\alp}_\mu
 \sbk{D_\alp,\bar{D}_{\dalp}}}V. 
\ee
Then its gauge transformation becomes simpler, 
\be
 X \to X+\Lmd+\bar{\Lmd}.  \label{X:gauge_trf}
\ee
Hence, a naive definition of a field strength superfield, 
\be
 \cW^{\rm naive}_\alp = -\frac{1}{4}\bar{D}^2D_\alp X, \label{cWnaive}
\ee
is gauge-invariant. 
However this does not transform correctly under the superconformal transformation. 
From (\ref{dsc_trf}), we see that 
$\dsc\cW_\alp^{\rm naive}$ contains $\bar{L}_{\dalp}$, 
which must be absent in the transformation of a chiral superfield. 
Thus we modify (\ref{cWnaive}) as 
\be
 \cW_\alp = -\frac{1}{4}\bar{D}^2\sbk{D_\alp X}_{\bE}, \label{def:cW}
\ee
where 
\be
 \sbk{D_\alp X}_{\bE} \equiv D_\alp X-\frac{1}{2}U^\mu\bar{\sgm}_\mu^{\dbt\bt}
 D_\alp D_\bt\bar{D}_{\dbt}X  \label{4D:sbk_bE}
\ee
is determined so that $\dsc\sbk{D_\alp X}_{\bE}$ 
does not contain $\bar{L}_{\dalp}$. 
In fact, this transforms correctly as a (superconformal) chiral multiplet 
with $w=\frac{3}{2}$, as shown in Appendix.~\ref{Trf:fss}. 
Notice that this modified superfield preserves the gauge invariance 
under (\ref{X:gauge_trf}). 
We have checked that each component of $\cW_\alp$ reproduces 
the correct forms of the field strength and the covariant derivative 
of the gaugino in Ref.~\cite{Sakamura:2011df}. 

Next we consider the non-Abelian case. 
In this case, there is no counterpart to $X$ in the Abelian case, and 
the gauge transformation is given by 
\be
 e^V \to \cU(e^\Lmd)e^V\cU(e^{\bar{\Lmd}}).  \label{gauge_trf:NA}
\ee
Thus a naive definition of the field strength superfield, 
\be
 \cW^{\rm naive}_\alp \equiv \frac{1}{4}\bar{D}^2\brkt{e^V D_\alp e^{-V}}, 
\ee
does not transform covariantly not only under the superconformal transformation, 
but also under the gauge transformation. 
We modify $\cW^{\rm naive}_\alp$ in the same strategy, and obtain 
\be
 \cW_\alp = \frac{1}{4}\bar{D}^2\sbk{e^V D_\alp e^{-V}}_{\bE},  \label{def:cW:NA}
\ee
where 
\bea
 \sbk{e^V D_\alp e^{-V}}_{\bE} \defa e^V D_\alp e^{-V}
 -\frac{1}{2}\bar{\sgm}^{\dbt\bt}_\mu D_\alp U^\mu
 \bar{D}_{\dbt}\brkt{e^V D_\bt e^{-V}} \nonumber\\
 &&+iD_\alp U^\mu e^V\der_\mu e^{-V}
 -iU^\mu\der_\mu\brkt{e^V D_\alp e^{-V}},  \label{4D:sbk_bE:NA}
\eea
is determined so that its $\dsc$-variation does not contain $\bar{L}_{\dalp}$. 
This transforms correctly as a (superconformal) chiral multiplet 
with the Weyl weight~$3/2$, as shown in Appendix~\ref{Trf:fss}. 
In fact, $\sbk{e^V D_\alp e^{-V}}_{\bE}$ transforms under the gauge 
transformation~(\ref{gauge_trf:NA}) as 
\be 
 \sbk{e^V D_\alp e^{-V}}_{\bE} \to 
 e^\Lmd\sbk{e^V D_\alp e^{-V}}_{\bE} e^{-\Lmd}+e^\Lmd D_\alp e^{-\Lmd}, 
\ee
so that $\cW_\alp$ transforms covariantly. 
\be
 \cW_\alp \to e^\Lmd\cW_\alp e^{-\Lmd}. 
\ee
Therefore, (\ref{def:cW:NA}) is the desired field strength superfield.

\subsection{Invariant action} \label{action_fml}
\subsubsection{\bdm{F}- and \bdm{D}-term formulae}
Now we construct invariant actions under the gauge and superconformal 
transformations. 
First, let us consider the chiral superspace integral of a chiral superfield~$W$, 
\be
 S_F[W] \equiv \int\dr^2\tht\;W+\hc. \label{S_F}
\ee
We can easily check that this is invariant 
under (\ref{dsc_trf}) when the Weyl weight of $W$ is 3. 
This is the superfield description of the $F$-term action formula 
in Ref.~\cite{Kugo:1982cu}. 

Unlike the chiral superspace integral, 
the full superspace integral of a real scalar superfield~$\Omg$ is not 
invariant by itself for any choice of the Weyl weight. 
An invariant action can be constructed with the aid of 
the gravitational superfield~$U^\mu$ as~\footnote{
The factor 2 is necessary to match the normalization of the $D$-term 
action formula in Ref.~\cite{Kugo:1982cu}. } 
\be
 S_D[\Omg] = 2\int\dr^4x\int\dr^4\tht\;\brkt{1+\frac{1}{3}E_1}\Omg, 
 \label{S_D}
\ee
where the Weyl weight of $\Omg$ is 2, and 
\be
 E_1 \equiv \frac{1}{4}\bar{\sgm}_\mu^{\dalp\alp}\sbk{D_\alp,\bar{D}_{\dalp}}U^\mu, 
\ee
which transforms as 
\be
 \dsc E_1 = -\frac{1}{2}\bar{D}^2D^\alp L_\alp
 +\frac{3i}{2}\sgm^\mu_{\alp\dalp}\der_\mu\bar{D}^{\dalp}L^\alp+\hc. 
\ee
The action~(\ref{S_D}) reproduces the $D$-term action formula 
in Ref.~\cite{Kugo:1982cu} up to linear order in the gravitational fields. 

In summary, 4D SUGRA action is described by using (\ref{S_F}) 
and (\ref{S_D}) as 
\bea 
 S^{\rm (4D)} \eql S_D\sbk{\Omg}
 +S_F\sbk{\Phi_C^3W-\frac{1}{4}\cW^\alp\cW_\alp}, \nonumber\\
 \Omg \eql -\frac{3}{2}\abs{\cU(\Phi_C)}^2e^{-K/3},  \label{expr:Omg:4D}
\eea
where $\Phi_C$ is a chiral compensator superfield with $w=1$, and 
$K$ and $W$ are the K\"{a}hler potential 
and the superpotential, which consist of only the physical superfields 
with $w=0$. 

Here we comment on the superconformal gauge-fixing. 
In order to obtain the usual Poincar\'{e} SUGRA from the superconformally 
symmetric action, we have to impose 
the gauge-fixing conditions to eliminate extra symmetries~$\bdm{D}$, 
$\bdm{S}$, $U(1)_A$, $\bdm{K}$. 
For example, they are given by 
\be
 C^\Omg = -\frac{3}{2}, \;\;\;\;\;
 \zeta^\Omg = 0, \;\;\;\;\;
 \arg(\phi_C) = 0, \;\;\;\;\;
 b_\mu = 0, \label{scGF_cond}
\ee
in the unit of the Planck mass, $M_{\rm Pl}=1$. 
Here $\sbk{C^\Omg,\zeta^\Omg,\cdots}$ is a real general multiplet 
corresponding to the real superfield~$\Omg$, 
and $\phi_C$ is the scalar component of 
the chiral multiplet corresponding to $\Phi_C$. 
The last condition is already imposed in our superfield description 
as mentioned in footnote~\ref{b_zero}. 
The gauge-fixing~(\ref{scGF_cond}) leads to the canonically normalized 
Einstein-Hilbert term. 
For a choice of $\Omg$ in (\ref{expr:Omg:4D}), 
the first and third conditions in (\ref{scGF_cond}) 
are summarized as $\phi_C=\exp(K|_0/6)$.

\subsubsection{Kinetic terms for $\bdm{U^\mu}$} \label{L_kin:U}
Since we have neglected terms beyond linear order in $U^\mu$, 
eqs.(\ref{S_F}) and (\ref{S_D}) do not contain its kinetic terms. 
In order to deal with them, we introduce quadratic terms 
in $U^\mu$ that are independent of the matter superfields. 
Namely, we extend (\ref{S_D}) as 
\be
 S_D[\Omg] = \int\dr^4x\int\dr^4\tht\;\brc{
 E_2+2\brkt{1+\frac{1}{3}E_1}\Omg}, \label{S_D2}
\ee
where $E_2$ is quadratic in $U^\mu$ and independent of the matter superfields. 
Since $\dsc E_2$ is linear in $U^\mu$, $E_2$ is determined 
by the requirement that a matter-independent part of $\dsc S_D[\Omg]$ 
vanishes up to {\it linear order} in the gravitational fields. 
We consider a case that $\Omg$ is given by (\ref{expr:Omg:4D}). 
In order to pick up the matter-independent part of $\Omg$, 
we expand the compensator superfield~$\Phi_C$ around 
the background value~$\vev{\phi_C}=\vev{\exp(K/6)}$ as 
\be
 \Phi_C = \vev{e^{K/6}}+\tl{\Phi}_C. 
\ee
Then, from (\ref{dsc_trf}), we have 
\bea
 \dsc\Omg \eql -\frac{3}{2}\vev{e^{K/6}}
 \brkt{\dsc\tl{\Phi}_C+iU^\mu\der_\mu\dsc\tl{\Phi}_C}\vev{e^{-K/3}}+\hc
 +\cdots \nonumber\\
 \eql \frac{1}{8}\brkt{\bar{D}^2D^\alp L_\alp+\hc}
 +\frac{1}{8}U^\mu\der_\mu\brkt{i\bar{D}^2D^\alp L_\alp+\hc}+\cdots.  
 \label{dsc:tlOmg}
\eea
Thus we can show that
\bea
 \dsc S_D[\Omg] \eql \int\dr^4x\int\dr^4\tht\;
 \brc{\dsc E_2
 +\frac{2}{3}\dsc E_1\Omg+2\brkt{1+\frac{1}{3}E_1}\dsc\Omg}.  
 \nonumber\\
 \eql \int\dr^4x\int\dr^4\tht\;\left\{
 \dsc E_2+\frac{1}{12}E_1\brkt{\bar{D}^2D^\alp L_\alp+\hc} \right. \nonumber\\ 
 &&\hspace{25mm}\left. 
 +\frac{1}{4}U^\mu\der_\mu\brkt{i\bar{D}^2D^\alp L_\alp+\hc}\right\}+\cdots 
 \nonumber\\
 \eql \int\dr^4x\int\dr^4\tht\;\brc{
 \dsc E_2-\frac{1}{6}E_1\brkt{\bar{D}^2D^\alp L_\alp+\hc}}+\cdots, 
 \label{dsc:S_D}
\eea
where the ellipses in (\ref{dsc:tlOmg}) and (\ref{dsc:S_D}) 
denote matter-dependent terms. 
We have performed partial integrals at the second equality. 
Therefore, $E_2$ must satisfy  
\be
 \dsc E_2 = \frac{1}{6}E_1\brkt{\bar{D}^2D^\alp L_\alp+\hc}.  \label{dsc:E_2}
\ee
From this condition, $E_2$ is identified as 
\be
 E_2 = -\frac{1}{8}U_\mu D^\alp\bar{D}^2D_\alp U^\mu
 +\frac{1}{3}E_1^2-\brkt{\der_\mu U^\mu}^2.  \label{def:E_2}
\ee


Let us compare (\ref{S_D2}) with the action 
in Refs.~\cite{Linch:2002wg,Buchbinder:1995uq}. 
The action~(\ref{S_D2}) is rewritten as 
\bea
 S_D[\Omg] \eql \int\dr^4x\int\dr^4\tht\;\left\{E_2
 -\frac{\vev{e^{-K/6}}}{4}\bar{\sgm}_\mu^{\dalp\alp}
 \sbk{D_\alp,\bar{D}_{\dalp}}U^\mu
 \brkt{\tl{\Phi}_C+\bar{\tl{\Phi}}_C}-3\abs{\tl{\Phi}_C}^2 \right. \nonumber\\
 &&\left.\hspace{25mm}
 -3i\vev{e^{-K/6}}U^\mu\der_\mu
 \brkt{\tl{\Phi}_C-\bar{\tl{\Phi}}_C}\right\}+\cdots \nonumber\\
 \eql \int\dr^4x\;\int\dr^4\tht\;\brc{E_2-3\abs{\tl{\Phi}_C}^2
 +2i\vev{e^{-K/6}}\der_\mu U^\mu\brkt{\tl{\Phi}_C-\bar{\tl{\Phi}}_C}}+\cdots.  
 \label{S_D3}
\eea
We have performed partial integrals. 
In Refs.~\cite{Linch:2002wg,Buchbinder:1995uq}, the compensator 
superfield~$\Sgm^{\rm cp}$ is fixed by the gauge-fixing as 
$\Sgm^{\rm cp}=\tl{e}_\mu^{\;\;\mu}+\tht^2F^\Sgm$ in our notation. 
From (\ref{sf:chiral}) and (\ref{def:cE}), on the other hand, we see that 
$\tl{\Phi}_C = \frac{1}{3}\vev{e^{K/6}}\tl{e}_\mu^{\;\;\mu}+\tht^2F^{\Phi_C}
+\cdots$ after the gauge-fixing. 
The ellipsis denotes terms involving phyiscal matter fields. 
Thus $\Sgm^{\rm cp}$ is identified as 
$\Sgm^{\rm cp} =3\vev{e^{-K/6}}\tl{\Phi}_C$. 
With this and (\ref{def:E_2}), we find that (\ref{S_D3}) agrees with 
(3.19) of Ref.~\cite{Linch:2002wg}.

\section{5D supergravity}  \label{5D_SUGRA}
Now we consider 5D SUGRA. 
We assume that the background geometry is flat, 
\be
 ds^2 = \eta_{\mu\nu}dx^\mu dx^\nu-dy^2.  \label{flat_metric}
\ee
We will extend it to the warped geometry in Sec.~\ref{warped_case}. 
The following superfield description is based on 
the superconformal formulation of Ref.~\cite{Kugo:2000af}-\cite{Kugo:2002js}. 

\subsection{Decomposition into $\bdm{\cN=1}$ multiplets} \label{decomp:sf}
The 5D superconformal transformations are devided into 
two parts~$\dscp$ and $\dscq$, where $\dscp$ forms an $\cN=1$ subalgebra, 
and $\dscq$ is the rest part. 
We mainly focus on $\dscp$ in the following discussion. 
We will consider $\dscq$ in Sec.~\ref{L_kin:5D}. 
As shown in Ref.~\cite{Kugo:2002js}, 
each 5D superconformal multiplet 
can be decomposed into $\cN=1$ superconformal multiplets, 
which only respect $\dscp$ manifestly. 

A hypermultiplet~$\mathbb H^a$ $(a=1,2,\cdots,n_C+n_H)$ 
is decomposed into two chiral multiplets~$(\Phi^{2a-1},\Phi^{2a})$, 
and a vector multiplet~$\mathbb V^I$ $(I=0,1,\cdots,n_V)$
is into $\cN=1$ vector and chiral multiplets~$(V^I,\Sgm^I)$. 
Here $n_C$ ($n_H$) and $n_V$ are the numbers of the compensator (physical) 
hypermultiplets and the physical vector multiplets. 
These $\cN=1$ multiplets are expressed by $\cN=1$ superfields 
as explained in the previous section. 
The explicit forms of these superfields are 
listed in Appendix~\ref{comp:superfields}. 
Here there is one point to notice. 
In the decompositions in Ref.~\cite{Kugo:2002js}, 
the $Z_2$-odd fields are dropped 
because such decompositions are considered only on the $S^1/Z_2$ boundaries 
where the $Z_2$-odd fields vanish, 
in order to describe couplings between the bulk multiplets 
and 4D multiplets localized on the boundaries. 
In this paper, on the other hand, 
the $\cN=1$ decompositions are considered in the bulk.  
Therefore, each component of the $\cN=1$ superfields 
listed in Appendix~\ref{comp:superfields} may be 
corrected by terms involving the $Z_2$-odd fields.  
We will come back to this point in Sec.~\ref{ID:Z2odd}. 

The 5D Weyl multiplet~${\mathbb E}_W$ (or the 5D gravitational multiplet)
is also decomposed into $\cN=1$ multiplets. 
Here ${\mathbb E}_W$ consists of 
the f\"unfbein~$e_M^{\;\;\underline{N}}$, the gravitini~$\psi_M^i$, 
the $\suU$ gauge field~$V_M^r$, the $\bdm{D}$ gauge field~$b_\mu$, 
an antisymmetric auxiliary tensor~$v^{\underline{M}\underline{N}}$, 
and other auxiliary fields. 
The 5D indices~$M,N=0,1,2,3,y$ and 
$\underline{M},\underline{N}=0,1,2,3,4$ denote 
the curved and the flat ones, and 
$i=1,2$ and $r=1,2,3$ denote the $\suU$-doublet 
and triplet indices. 
In the case that the extra dimension is compactified on $S^1/Z_2$, 
the $Z_2$-even part of $\bE_W$ forms the $\cN=1$ Weyl multiplet 
and a real general multiplet, which can be expressed by 
the following superfields.\footnote{
Just like in the 4D SUGRA case, 
the $\bdm{D}$ gauge field~$b_\mu$ can be set to zero, which corresponds to 
the $\bdm{K}$ gauge fixing.  }
\bea
 U^\mu \eql \brkt{\tht\sgm^\nu\bar{\tht}}\tl{e}_\nu^{\;\;\mu}
 +i\bar{\tht}^2\brkt{\tht\sgm^\nu\bar{\sgm}^\mu\psi_\nu^+}
 -i\tht^2\brkt{\bar{\tht}\bar{\sgm}^\nu\sgm^\mu\bar{\psi}^+_\nu}
 +\tht^2\bar{\tht}^2\brkt{V^{3\mu}+v^{\mu}_{\;\;4}
 -\frac{1}{4}\ep^{\mu\nu\rho\tau}\der_\nu\tl{e}_{\rho\tau}},  \nonumber\\
 V_E \eql \brc{1-\frac{1}{6}\brkt{\cE+\bar{\cE}}}
 \left\{\brkt{1+\tl{e}_y^{\;\;4}}+2\tht\psi_y^-+2\bar{\tht}\bar{\psi}_y^-
 -i\tht^2\brkt{V_y^1+iV_y^2}+i\bar{\tht}^2\brkt{V_y^1-iV_y^2} 
 \right. \nonumber\\
 &&\left. \hspace{30mm}
 -\frac{2}{3}\brkt{\tht\sgm^\mu\bar{\tht}}
 \brc{V_\mu^3-2v_{\mu 4}}+\cdots\right\},  \label{def:UV_E}
\eea
where $\tl{e}_M^{\;\;N}\equiv e_M^{\;\;\underline{N}}-\dlt_M^{\;\;N}$, 
the 2-component spinors~$\psi_M^\pm$ are defined 
from the 4-component notation of Ref.~\cite{Kugo:2002js} 
through (\ref{rel:2-4spinor}), and 
\be
 \cE \equiv \tl{e}_\mu^{\;\;\mu}-2i\tht\sgm^\mu\bar{\psi}^+_\mu. 
\ee 
As mentioned in Sec.~\ref{def:sf:4D}, we need not discriminate 
the curved and the flat indices since we keep the gravitational fields 
up to linear order and the background geometry is flat. 
Thus we again omit the underlines of the flat indices. 

The $\dscp$-transformation laws of the above superfields are 
the same as the $\dsc$-transformation in the previous section. 
Namely, 
\bea
 \dscp U^\mu \eql \frac{1}{2}\sgm^\mu_{\alp\dalp}
 \brkt{\bar{D}^{\dalp}L^\alp-D^\alp\bar{L}^{\dalp}}, \nonumber\\
 \dscp V_E \eql \brkt{-\frac{1}{4}\bar{D}^2L^\alp D_\alp
 -\frac{i}{2}\sgm^\mu_{\alp\dalp}\bar{D}^{\dalp}L^\alp\der_\mu
 +\frac{1}{24}\bar{D}^2D^\alp L_\alp+\hc}V_E, \nonumber\\
 \dscp\Phi^{\oa} \eql \brkt{-\frac{1}{4}\bar{D}^2L^\alp D_\alp
 -i\sgm^\mu_{\alp\dalp}\bar{D}^{\dalp}L^\alp\der_\mu
 -\frac{1}{8}\bar{D}^2D^\alp L_\alp}\Phi^{\oa}, \nonumber\\
 \dscp V^I \eql \brkt{-\frac{1}{4}\bar{D}^2L^\alp D_\alp
 -\frac{i}{2}\sgm^\mu_{\alp\dalp}\bar{D}^{\dalp}L^\alp\der_\mu+\hc}V^I, 
 \nonumber\\
 \dscp\Sgm^I \eql \brkt{-\frac{1}{4}\bar{D}^2L^\alp D_\alp
 -i\sgm^\mu_{\alp\dalp}\bar{D}^{\dalp}L^\alp\der_\mu}\Sgm^I,  
 \label{dscp:sf}
\eea
where the index~$\oa$ runs over the whole $2(n_C+n_H)$ chiral multiplets 
coming from the hypermultiplets. 
Notice that the Weyl weights of these superfields are 
$w(V_E)=-1$, $w(\Phi^{\oa})=3/2$, and $w(V^I)=w(\Sgm^I)=0$. 
Since $V_E$ has a nonzero background value~$\vev{V_E}=1$, 
the gravitational fluctuation superfield is $\tl{V}_E\equiv V_E-1$, 
and its transformation law should be written as 
\be
 \dscp\tl{V}_E = \frac{1}{24}\brkt{\bar{D}^2D^\alp L_\alp+\hc}, 
\ee
up to the order concerned in this paper. 
However, since $V_E$ behaves as a real general multiplet under $\dscp$, 
it is sometimes convenient to treat it as a matter multiplet.

\subsection{Superfield action without gravitational fluctuation modes}
In order to identify the gravitational coupling to the matter superfields, 
we start with the 5D superfield action 
in which the gravitational fields are fixed 
to their background values. 
Such an action was derived in Ref.~\cite{Abe:2004ar,Paccetti:2004ri} as~\footnote{
The sign of the $d^2\tht$-integral in $\cL_{\rm hyper}$ is opposite 
to that of Ref.~\cite{Abe:2004ar}. 
This stems from the sign difference in the definitions of the $F$-terms 
in the chiral superfields there. 
} 
\bea
 S_0 \eql \int\dr^5x\;\brkt{\cL_0^{\rm hyper}+\cL_0^{\rm vector}}, \nonumber\\
 \cL_0^{\rm hyper} \eql -2\int\dr^4\tht\;
 d_{\oa}^{\;\;\ob}\bar{\Phi}_{\ob}\brkt{e^{-2ig V^It_I}}^{\oa}_{\;\;\oc}\Phi^{\oc} 
 \nonumber\\
 &&+\int\dr^2\tht\;\Phi^{\oa}d_{\oa}^{\;\;\ob}\rho_{\ob\oc}
 \brkt{\der_y-2ig\Sgm^It_I}^{\oc}_{\;\;\od}\Phi^{\od}+\hc, \nonumber\\
 \cL_0^{\rm vector} \eql -\int\dr^4\tht\;
 C_{IJK}\cV_0^I\cV_0^J\cV_0^K  \nonumber\\
 &&+\int\dr^2\tht\;\frac{3C_{IJK}}{2}\brc{
 -\Sgm^I\cW_0^J\cW_0^K
 +\frac{1}{12}\bar{D}^2\brkt{\cZ_0^{IJ\alp}}\cW_{0\alp}^K}+\hc,  \label{L_0^vector}
\eea
where $\bar{\Phi}_{\ob}\equiv (\Phi^{\ob})^\dagger$, 
$d_{\oa}^{\;\;\ob}\equiv\diag(\id_{2n_C},-\id_{2n_H})$, 
$\rho_{\oa\ob}\equiv i\sgm_2\otimes\id_{n_C+n_H}$, 
and real constants~$C_{IJK}$ are a completely symmetric tensor. 
The generators~$t_I$ ($I=0,1,\cdots,n_V$) are anti-hermitian 
and normalized as $\tr(t_It_J)=-\frac{1}{2}$. 
The gauge couplings~$g$ can take different values for each simple 
or Abelian factor of the gauge group. 
The assumption of the 5D flat geometry~(\ref{flat_metric}) 
is equivalent to that 
the compensator hypermultiplets~$\bH^a$ ($a=1,\cdots,n_C$) are gauge-singlets. 
For simplicity, we consider a case of Abelian gauge group in the following. 
An extension to the non-Abelian case will be discussed 
in Sec.~\ref{NAcase}. 
Then the field strength superfields~$\cW_{0\alp}^I$ and $\cV_0^I$ are defined as 
\bea
 \cW_{0\alp}^I \defa -\frac{1}{4}\bar{D}^2D_\alp V^I, \nonumber\\
 \cV_0^I \defa -\der_y V^I+\Sgm^I+\bar{\Sgm}^I.   \label{def:cWV0}
\eea
Here and henceforth, the suffix 0 denotes quantities 
that the gravitational fluctuation modes are dropped. 
The second line in $\cL_0^{\rm vector}$ corresponds to the Chern-Simons terms, 
and $\cZ_0^{IJ\alp}$ is defined as~\cite{ArkaniHamed:2001tb} 
\be
 \cZ^{IJ\alp}_0 \equiv V^I D^\alp\der_y V^J-\der_y V^I D^\alp V^J. 
\ee
The gauge kinetic terms arise from the first term of the second line 
in $\cL_0^{\rm vector}$ after the superconformal gauge-fixing.

\section{Gravitational couplings} \label{grav_cp}
Now we turn on the gravitational fluctuation fields, and specify 
their couplings to the matter superfields. 
The easiest one to obtain is couplings with $\tl{V}_E=V_E-1$. 
In our previous work~\cite{Abe:2004ar}, 
we have already found the $V_E$-dependence of the action. 
It only appears in the $d^4\tht$-integral as 
\bea
 S \eql -\int\dr^5x\int\dr^4\tht\;V_E\brc{
 2d_{\oa}^{\;\;\ob}\bar{\Phi}_{\ob}\brkt{e^{-2igV^It_I}}^{\oa}_{\;\;\oc}
 \Phi^{\oc}+C_{IJK}V_E^{-3}\cV_0^I\cV_0^J\cV_0^K}+\cdots.  \label{V_E-dependence}
\eea

Next we specify couplings with the other gravitational superfields. 
We devide the 5D action into the sectors of 
the hypermultiplets, the vector multiplets, and 
the gravitational kinetic terms.

\subsection{Hypermultiplet sector}
Let us first consider the hypermultiplet sector. 
We can easily specify the couplings with $U^\mu$ defined in (\ref{def:UV_E})
by the procedure explained in Sec.~\ref{4D_SUGRA}. 
Namely, replace (anti-)chiral superfields in the $d^4\tht$-integral 
with the embedded ones defined in (\ref{def:cU}), and modify 
the integrand as (\ref{S_D2}). 
Here $E_1$ and $E_2$ are constructed from $U^\mu$ in (\ref{def:UV_E}). 

However these procedures are not enough to construct 
an invariant action under the $\dscp$-transformation. 
Note that $\der_y\Phi^{\oa}$ does not transform as a chiral multiplet 
under~$\dscp$, 
\bea
 \dscp\der_y\Phi^{\oa} \eql \der_y\brkt{\dscp\Phi^{\oa}} \nonumber\\
 \eql \brkt{-\frac{1}{4}\bar{D}^2L^\alp D_\alp
 -i\sgm^\mu_{\alp\dalp}\bar{D}^{\dalp}L^\alp\der_\mu
 -\frac{1}{8}\bar{D}^2D^\alp D^\alp L_\alp}\der_y\Phi^{\oa} \nonumber\\
 &&+\brkt{-\frac{1}{4}\bar{D}^2\der_y L^\alp D_\alp
 -i\sgm^\mu_{\alp\dalp}\bar{D}^{\dalp}\der_y L^\alp\der_\mu
 -\frac{1}{8}\bar{D}^2D^\alp\der_y L_\alp}\Phi^{\oa}. 
\eea
In order to eliminate 
the terms involving $\der_y L_\alp$ in the second line, 
we introduce a spinor superfield~$\Psi_\alp$ that transforms as 
\be
 \dscp\Psi_\alp = -\der_y L_\alp,  \label{dscp:Psi}
\ee
and covariantize the derivative~$\der_y$ as 
\be
 \hat{\der}_y \equiv \der_y-\frac{1}{4}\bar{D}^2\Psi^\alp D_\alp
 -i\sgm^\mu_{\alp\dalp}\bar{D}^{\dalp}\Psi^\alp\der_\mu
 -\frac{w}{12}\bar{D}^2D^\alp\Psi_\alp,  \label{def:hatder_y1}
\ee
where $w$ is the Weyl weight of a superfield which $\hat{\der}_y$ acts on. 
Then, $\hat{\der}_y\Phi^{\oa}$ transforms just in the same way as 
$\Phi^{\oa}$ does. 
As a result, $\cL_0^{\rm hyper}$ is promoted as 
\bea
 \cL^{\rm hyper} \eql 2\int\dr^4\tht\;
 \brc{\frac{\Omg^{\rm hyper}_{\rm c}}{3}E_2
 +\brkt{1+\frac{1}{3}E_1}\Omg^{\rm hyper}}  \nonumber\\
 &&+\int\dr^2\tht\;\Phi^{\oa}d_{\oa}^{\;\;\ob}\rho_{\ob\oc}
 \brkt{\hat{\der}_y-2ig\Sgm^I t_I}^{\oc}_{\;\;\od}\Phi^{\od}+\hc, 
 \label{L^hyper}
\eea
where 
\be
 \Omg^{\rm hyper} \equiv -V_E d_{\oa}^{\;\;\ob}
 \cU(\bar{\Phi}_{\ob})\brkt{e^{-2ig V^I t_I}}^{\oa}_{\;\;\oc}
 \cU(\Phi^{\oc}), 
\ee
and $\Omg^{\rm hyper}_{\rm c}$ is a constant part of $\Omg^{\rm hyper}$, 
which is determined by the $\bdm{D}$ gauge fixing condition.

\subsection{Vector multiplet sector} \label{vector_sector}
Next we consider the vector multiplet sector. 
The gauge transformation~$\dgt$ is given by  
\bea
 \dgt V^I \eql \cU(\Lmd^I)+\cU(\bar{\Lmd}^I), \nonumber\\
 \dgt\Sgm^I \eql \hat{\der}_y\Lmd^I, \nonumber\\
 \dgt\Phi^{\oa} \eql 2ig\Lmd^I(t_I)^{\oa}_{\;\;\ob}\Phi^{\ob},  
 \label{dgt:superfield}
\eea
where the transformation parameter~$\Lmd^I$ is a chiral superfield. 
We can easily check that (\ref{L^hyper}) is invariant under this transformation. 

The field strength superfields~$\cW^I_{0\alp}$ should be modified 
as (\ref{def:cW}) with the aid of $U^\mu$. 
\bea
 \cW^I_\alp \defa -\frac{1}{4}\bar{D}^2
 \brc{D_\alp X^I-\frac{1}{2}U^\mu\bar{\sgm}_\mu^{\dbt\bt}
 D_\alp D_\bt\bar{D}_{\dbt}X^I}, \nonumber\\
 X^I \defa \brkt{1+\frac{1}{4}U^\mu\bar{\sgm}_\mu^{\dalp\alp}
 \sbk{D_\alp,\bar{D}_{\dalp}}}V^I.  \label{def:cW^I}
\eea
This is invariant under $\dgt$. 

Now we modify the other field strength superfield~$\cV_0^I$. 
First, $\Sgm^I$ and $\bar{\Sgm}^I$ in (\ref{def:cWV0}) should be replaced with
$\cU(\Sgm^I)$ and $\cU(\bar{\Sgm}^I)$. 
Thus the first term in (\ref{def:cWV0}) must be modified so that 
its gauge transformation is $\cU(\hat{\der}_y\Lmd^I+\hat{\der}_y\bar{\Lmd}^I)$. 
Here we redefine $\hat{\der}_y$ as 
\be
 \hat{\der}_y = \der_y-\brkt{\frac{1}{4}\bar{D}^2\Psi^\alp D_\alp
 +\frac{1}{2}\bar{D}^{\dalp}\Psi^\alp\bar{D}_{\dalp}D_\alp
 +\frac{w+n}{24}\bar{D}^2D^\alp\Psi_\alp+\hc}, 
\ee
where $n$ is a chiral weight, \ie, the charge of $U(1)_A\subset\suU$, and $(w+n)^\dagger=w-n$.  
This reduces to the definition~(\ref{def:hatder_y1}) 
when it acts on a chiral superfield. 
Since $X^I$ transforms as $\dgt X^I=\Lmd^I+\bar{\Lmd}^I$, 
we find that $\dgt(\hat{\der}_y X^I)=\hat{\der}_y\Lmd^I+\hat{\der}_y\bar{\Lmd}^I$. 
Therefore, $\cV_0^I$ is modified as 
\bea
 \cV^I \defa \brkt{1-\frac{1}{4}U^\mu\bar{\sgm}^{\dalp\alp}_\mu
 \sbk{D_\alp,\bar{D}_{\dalp}}}
 \brkt{-\hat{\der}_y X^I+\Sgm^I+\bar{\Sgm}^I}  \nonumber\\
 \eql -\brkt{\hat{\der}_y+\frac{1}{4}\der_y U^\mu
 \bar{\sgm}^{\dalp\alp}_\mu\sbk{D_\alp,\bar{D}_{\dalp}}}V^I
 +\cU(\Sgm^I)+\cU(\bar{\Sgm}^I). 
\eea 
This transforms under $\dscp$ in the same way as $V^I$ does. 

Hence the $d^4\tht$-integral in $\cL^{\rm vector}_0$ is promoted as 
\be
 \cL^{\rm vector} = 2\int\dr^4\tht\;
 \brc{\frac{\Omg^{\rm vector}_{\rm c}}{3}E_2
 +\brkt{1+\frac{1}{3}E_1}\Omg^{\rm vector}}
 +\brc{\int\dr^2\tht\;W^{\rm CS}+\hc}, 
 \label{cL^vector1}
\ee
where 
\be
 \Omg^{\rm vector} = -V_E^{-2}\frac{C_{IJK}}{2}\cV^I\cV^J\cV^K, 
 \label{expr:Omg^vector}
\ee
and $\Omg^{\rm vector}_{\rm c}$ is a constant part of $\Omg^{\rm vector}$, 
which is determined by the $\bdm{D}$ gauge fixing condition. 
The holomorphic function~$W^{\rm CS}$ 
includes the Chern-Simons terms, and will be specified in the following. 

Notice that there is no operation in the superconformal tensor calculus 
corresponding to $D_\alp$ ($\bar{D}_{\dalp}$). 
Hence promoting $\bar{D}^2\cZ^{IJ}_{0\alp}$ in (\ref{L_0^vector})
to SUGRA is a nontrivial task. 
We have to modify it so that it transforms 
just in the same way as $\cW^I_\alp$ under $\dscp$, 
\ie, as a spinor chiral superfield with $w=\frac{3}{2}$.  
The strategy is similar to that we applied in Sec.~\ref{fss}.
Since $\hat{\der}_y X^I$ transforms in the same way as $X^I$, 
\be
  \sbk{D_\alp\der_y X^I}_{\bE} \equiv D_\alp\hat{\der}_y X^I
 -\frac{1}{2}\bar{\sgm}^{\dbt\bt}_\mu U^\mu D_\alp D_\bt\bar{D}_{\dbt}\der_y X^I 
 \label{def:sbk_bE1}
\ee
also transforms in the same way as $\sbk{D_\alp X^I}_{\bE}$ 
defined in (\ref{4D:sbk_bE}).  
Namely, their transformation laws are given by 
\bea
 \dscp\sbk{\cY^I_\alp}_{\bE} 
 \eql \brkt{-\frac{1}{4}\bar{D}^2L^\bt D_\bt
 -i\sgm^\mu_{\bt\dbt}\bar{D}^{\dbt}L^\bt\der_\mu}\cY_\alp^I \nonumber\\
 &&+\brkt{-\frac{1}{4}D_\alp\bar{D}^2 L^\bt
 -\frac{1}{2}D_\alp\bar{D}^{\dbt}L^\bt\bar{D}_{\dbt}}\cY_\bt^I, 
 \label{dscp:cY}
\eea
where $\cY^I_\alp=D_\alp X^I,D_\alp\der_y X^I$. 
Furthermore, we define 
\be
 \sbk{\cX^I\cY^J_\alp}_{\bE}
 \equiv \cX^I\sbk{\cY_\alp^J}_{\bE}-\frac{1}{2}\bar{\sgm}_\mu^{\dbt\bt}
 \brkt{U^\mu D_\bt\bar{D}_{\dbt}\cX^I\cY^J_\alp
 +D_\alp U^\mu\bar{D}_{\dbt}\cX^I\cY^J_\bt}, 
\ee
where $\cX^I=X^I,\hat{\der}_y X^I$, 
so that the $\bar{L}$-dependent terms in its $\dscp$-variation 
are cancelled. 
Then we find that this also follow the same transformation law 
as (\ref{dscp:cY}). 
This indicates that $\bar{D}^2\sbk{\cX^I\cY^J_\alp}_{\bE}$ transforms 
under $\dscp$ just in the same way as $\cW_\alp^I$ does. 

However, (\ref{def:sbk_bE1}) is not a unique way to promote $D_\alp\der_y X^I$. 
The following quantity also transforms as (\ref{dscp:cY}). 
\bea
  \sbk{\der_y D_\alp X^I}_{\bE} \defa \der_y\sbk{D_\alp X^I}_{\bE}
 +\brkt{-\frac{1}{4}\bar{D}^2\Psi^\bt D_\bt
 -i\sgm^\mu_{\bt\dbt}\bar{D}^{\dbt}\Psi^\bt\der_\mu}D_\alp X^I 
 \nonumber\\
 &&+\brkt{-\frac{1}{4}D_\alp\bar{D}^2\Psi^\bt
 -\frac{1}{2}D_\alp\bar{D}^{\dbt}\Psi^\bt\bar{D}_{\dbt}}D_\bt X^I.   
 \label{def:sbk_bE2}
\eea
The variations of the second and the third terms cancel 
the $\der_y L$-dependent terms 
coming from the variation of the first term. 
The quantities in (\ref{def:sbk_bE1}) and (\ref{def:sbk_bE2}) are related 
to each other as   
\be
 \sbk{\der_y D_\alp X^I}_{\bE} = \sbk{D_\alp\der_y X^I}_{\bE}
 +\frac{1}{4}\brkt{\sgm^\mu_{\alp\dbt}\der_y U_\mu
 +\bar{D}_{\dbt}\Psi_\alp-D_\alp\bar{\Psi}_{\dbt}}D^2\bar{D}^{\dbt}X^I.  
 \label{rel:DderXs}
\ee
Therefore, the promoted form of $\cZ_{0\alp}^{IJ}$ is expressed as 
\be
 \cZ_\alp^{IJ} = (1-a)\sbk{X^ID_\alp\der_y X^J}_{\bE}
 +a\sbk{X^I\der_y D_\alp X^J}_{\bE} 
 -\sbk{\hat{\der}_y X^ID_\alp X^J}_{\bE},  
 \label{expr:cZ^IJ}
\ee
where $a$ is a constant that cannot be determined only 
by the $\dscp$-invariance of the action. 
As shown in Appendix~\ref{GI:L_CS}, we find that 
the gauge invariance requires that $a=1$. 
As a result, $W^{\rm CS}$ is expressed as 
\bea
 W^{\rm CS} \eql \frac{3C_{IJK}}{2}\brc{-\Sgm^I\cW^J\cW^K
 +\frac{1}{12}\bar{D}^2\brkt{\cZ^{IJ\alp}}\cW^K_\alp}, \nonumber\\
 \cZ^{IJ}_\alp \eql \sbk{X^I\der_y D_\alp X^J}_{\bE}
 -\sbk{\hat{\der}_y X^I D_\alp X^J}_{\bE}.   \label{expr:W^CS}
\eea

\subsection{Non-Abelian case} \label{NAcase}
Here we consider the case of a non-Abelian gauge group. 
For simplicity, we assume that the whole gauge group is a simple 
non-Abelian group.  
In this case, it is convenient to use a matrix 
notation~$(\Sgm,V)\equiv 2ig(\Sgm^I,V^I)t_I$. 
The Lagrangian of hypermultiplet sector is written as 
\bea
 \cL^{\rm hyper} \eql 2\int\dr^4\tht\;
 \brc{\frac{\Omg^{\rm hyper}_{\rm c}}{3}E_2
 +\brkt{1+\frac{1}{3}E_1}\Omg^{\rm hyper}} 
 +\sbk{\int\dr^2\tht\;W^{\rm hyper}+\hc}, \nonumber\\
 \Omg^{\rm hyper} \defa -V_E\brc{
 \cU(\Phi_{\rm odd}^\dagger)\tl{d}\brkt{e^{V}}^t\cU(\Phi_{\rm odd})
 +\cU(\Phi_{\rm even}^\dagger)\tl{d}e^{-V}\cU(\Phi_{\rm even})}, \nonumber\\
 W^{\rm hyper} \defa \Phi_{\rm odd}^t\tl{d}\brkt{\hat{\der}_y-\Sgm}\Phi_{\rm even}
 -\Phi_{\rm even}^t\tl{d}\brkt{\hat{\der}_y+\Sgm^t}\Phi_{\rm odd}. 
 \label{cL^hyper:NA}
\eea
where $\tl{d}=\diag(\id_{n_C},-\id_{n_H})$, 
and $\Phi_{\rm odd}$ and $\Phi_{\rm even}$ are $(n_C+n_H)$-dimensional 
column vectors that consist of $\Phi^{2a-1}$ and $\Phi^{2a}$, respectively. 
This Lagrangian is invariant under the following gauge transformation. 
\bea
 &&e^V \to e^{\cU(\Lmd)}e^V e^{\cU(\Lmd^\dagger)}, \;\;\;\;\;
 \Sgm \to e^\Lmd\brkt{\Sgm-\hat{\der}_y}e^{-\Lmd}, \nonumber\\
 &&\Phi_{\rm odd} \to \brkt{e^{-\Lmd}}^t\Phi_{\rm odd}, \;\;\;\;\;
 \Phi_{\rm even} \to e^\Lmd\Phi_{\rm even},  \label{5Dgauge_trf:NA}
\eea
if $e^\Lmd$ commutes with $\tl{d}$. 

Next we consider the vector multiplet sector. 
The index~$I$ is now understood as that of the adjoint representation of 
the gauge group. 
For the constant tensor~$C_{IJK}$, there is a set of 
hermitian matrices~$\brc{T_I}$, which satisfies~\cite{Kugo:2000af}
\be
 C_{IJK} = \frac{1}{6}\tr\brkt{T_I\brc{T_J,T_K}}. 
\ee
In general, $\brc{T_I}$ are not normalized, 
and related to the normalized anti-hermitian matrices~$\brc{t_I}$ 
through $t_I=iT_I/c$, where $c$ is a real constant.\footnote{
In a case that the gauge group is a product of simple groups 
and Abelian groups, the constant~$c$ can take different values 
for each simple or Abelian factor group. }
Then, an extension of $\Omg^{\rm vector}$ in (\ref{cL^vector1}) 
to the non-Abelian gauge group is given by 
\be
 \Omg^{\rm vector} \equiv \frac{c^3}{48g^3}V_E^{-2}\tr\brkt{\cV^3}, 
 \label{def:Omg^vector:NA}
\ee
where
\be
 \cV \equiv -\brkt{\hat{\der}_y+\frac{1}{4}\der_y U^\mu
 \bar{\sgm}_\mu^{\dalp\alp}\sbk{D_\alp,\bar{D}_{\dalp}}} e^V e^{-V} 
 +\cU(\Sgm)+e^V\cU(\Sgm^\dagger)e^{-V}. 
\ee
The $U^\mu$-, $\Psi_\alp$($\bar{\Psi}_{\dalp}$)-independent part 
of $W^{\rm CS}$ in (\ref{cL^vector1}) is expressed as~\cite{Hebecker:2008rk}  
\bea
 W^{\rm CS}_0 \eql -\frac{c^3}{16g^3}\tr\sbk{
 -\Sgm\cW_0^2+\frac{1}{24}\bar{D}^2\brkt{\cZ_0^\alp}
 \brkt{\cW_{0\alp}-\frac{1}{4}\cW_{0\alp}^{(2)}}},  \label{NA:W^CS_0}
\eea
Here we have taken the Wess-Zumino gauge, and thus,  
\bea
 \cW_{0\alp} \defa \frac{1}{4}\bar{D}^2\brkt{e^{V}D_\alp e^{-V}}
 = -\frac{1}{4}\bar{D}^2D_\alp V
 +\frac{1}{8}\bar{D}^2\sbk{D_\alp V,V},  \nonumber\\ 
 \cZ_0^\alp \defa \brc{V,\der_y D^\alp V}-\brc{\der_y V,D^\alp V},  
\eea
and $\cW_{0\alp}^{(2)}$ is a quadratic part of $\cW_{0\alp}$, \ie, 
$\cW^{(2)}_{0\alp} \equiv \frac{1}{8}\bar{D}^2\sbk{D_\alp V,V}$. 
The couplings to the gravitational superfields~$U^\mu$ 
and $\Psi_\alp$ ($\bar{\Psi}_{\dalp}$) can be obtained 
in the same manner as in the Abelian case. 
Namely, (\ref{NA:W^CS_0}) is modified as 
\be
 W^{\rm CS} = -\frac{c^3}{16g^3}\tr\sbk{
 -\Sgm\cW^2+\frac{1}{24}\bar{D}^2\brkt{\cZ^\alp}
 \brkt{\cW_\alp-\frac{1}{4}\cW^{(2)}_\alp}},  \label{expr:W^CS:NA}
\ee
where $\cW_\alp$ is defined in (\ref{def:cW:NA}), 
$\cW_\alp^{(2)}$ is its quadratic part in $V$, 
and 
\be
 \cZ_\alp \equiv \sbk{\brc{X,\der_y D_\alp X}}_{\bE}
 -\sbk{\brc{\der_y X,D_\alp X}}_{\bE}. 
\ee
Here $X\equiv \brkt{1+\frac{1}{4}U^\mu\bar{\sgm}_\mu^{\dalp\alp}
\sbk{D_\alp,\bar{D}_{\dalp}}}V$, and the definition of 
$\sbk{\cdots}_{\bE}$ is similar to the Abelian case. 
We can see that (\ref{expr:W^CS:NA}) reduces to (\ref{expr:W^CS}) 
when the gauge group is Abelian. 
However, the gauge-invariance of the action is not manifest now
since we have chosen the Wess-Zumino gauge.

\ignore{
In this case, it is convenient to use a matrix notation~$A_M\equiv A_M^It_I$, 
$F_{MN}\equiv F_{MN}^It_I$, $\cdots$. 
The 5D Chern-Simons terms are written as~\cite{Kugo:2000af}
\be
 \cL^{\rm CS} = \frac{C_{IJK}}{8}\ep^{MNLPQ}A_M^I
 \left(F^J_{NL}F^K_{PQ}+\frac{g}{2}\sbk{A_N,A_L}^JF^K_{PQ}
 +\frac{g^2}{10}\sbk{A_N,A_L}^J\sbk{A_P,A_Q}^K \right), 
 \label{L_CS^bs}
\ee
where $F^I_{MN}\equiv\der_M A_N^I-\der_N A_M^I-g\sbk{A_M,A_N}^I$, 
and $\sbk{A_M,A_N}^I\equiv -2\tr(t_I\sbk{A_M,A_N})$. 
The anti-hermitian matrices~$t_I$ is normalized as 
$\tr(t_It_J)=-\frac{1}{2}\dlt_{IJ}$. 
Then, (\ref{L_CS^bs}) is rewritten as
\bea
 \cL^{\rm CS} \eql \frac{ic^3}{6}\ep^{MNLPQ}\tr\brkt{
 \frac{1}{4}A_M F_{NL}F_{PQ}+\frac{g}{4}A_MA_NA_LF_{PQ}
 +\frac{g^2}{10}A_MA_NA_LA_PA_Q} \nonumber\\
 \eql \frac{ic^3}{6}\ep^{\mu\nu\rho\sgm}\tr\sbk{
 \frac{3}{4}A_y F_{\mu\nu}F_{\rho\sgm}
 -\frac{1}{2}\brc{A_\mu,\der_y A_\nu}F_{\rho\sgm}
 -\frac{g}{4}\brc{A_\mu,\der_y A_\nu}A_\rho A_\sgm}. 
\eea
}

\subsection{Kinetic terms for gravitational superfields} \label{L_kin:5D}
Now we consider the kinetic terms for the gravitational superfields. 
Those for $U^\mu$ are obtained just in the same way 
as we did in Sec.~\ref{L_kin:U}. 
In addition to them, we can construct the following invariant Lagrangian term 
from (\ref{dscp:sf}) and (\ref{dscp:Psi}).  
\bea
 \cL_{\cC} \eql 2\int\dr^4\tht\;b\,\cC^\mu\cC_\mu, \nonumber\\
 \cC_\mu \defa \der_y U_\mu+\frac{1}{2}\bar{\sgm}_\mu^{\dalp\alp}
 \brkt{\bar{D}_{\dalp}\Psi_\alp-D_\alp\bar{\Psi}_{\dalp}},  
 \label{cL_cC}
\eea
where $b$ is a real constant. 
Since this is invariant both under $\dscp$ and $\dgt$ by itself, 
a constant~$b$ cannot be determined by those symmetries. 
Hence we now consider the invariance 
under the rest part of the 5D superconformal transformation~$\dscq$. 
This contains the translation along the fifth dimension~$\bdm{P^4}$ 
and the Lorentz transformation~$\bdm{M_{\mu 4}}$ that mixes $x^\mu$ and $y$. 
Comparing our gravitational superfields with 
the counterparts in Ref.~\cite{Linch:2002wg}, 
we expect that the transformation parameters of $\dscq$ form 
a chiral superfield~$Y=\xi^4+\cdots$, 
and a real scalar superfield~$N=\brkt{\tht\sgm^\mu\bar{\tht}}\lmd_{\mu 4}+\cdots$, 
where $\xi^4$ and $\lmd_{\mu 4}$ are the transformation parameters 
for $\bdm{P^4}$ and $\bdm{M_{\mu 4}}$, respectively. 
This will be confirmed in the next subsection. 
There, we will also see that $\Psi_\alp$ contains $e_y^{\;\;\underline{\mu}}$. 
Thus there should be another gravitational superfield that contains $e_\mu^{\;\;4}$. 
From the correspondence to Ref.~\cite{Linch:2002wg}, 
such additional superfield is expected to be a real scalar superfield~$U^4$ 
whose superconformal transformations are given by $\dscp U^4=0$ and $\dscq U^4=N$.  

We find the $\dscq$-transformation laws of the $\cN=1$ superfields as 
\bea
 \dscq U^\mu \eql 0, \;\;\;\;
 \dscq\tl{V}_E = \frac{1}{2}\der_y\brkt{Y+\bar{Y}}, \;\;\;\;
 \dscq\Psi_\alp = \frac{i}{2}D_\alp\tl{N}, \;\;\;\;
 \dscq U^4 = N, \nonumber\\
 \dscq\Phi_{\rm odd} \eql Y\der_y\Phi_{\rm odd}
 -\frac{i}{4}\bar{D}^2\brc{\tl{N}\brkt{e^{-V}}^t\bar{\Phi}_{\rm even}}, 
 \nonumber\\
 \dscq\Phi_{\rm even} \eql Y\der_y\Phi_{\rm even}
 +\frac{i}{4}\bar{D}^2\brc{\tl{N}e^V\bar{\Phi}_{\rm odd}}, 
 \nonumber\\
 \dscq e^V \eql \frac{1}{2}\brkt{Y+\bar{Y}}\der_y e^V
 +i\tl{N}\brkt{\Sgm e^V-e^V\Sgm^\dagger}, \nonumber\\
 \dscq\Sgm \eql \der_y\brkt{Y\Sgm}
 -\frac{i}{8}\bar{D}^2\brkt{D^\alp\tl{N}D_\alp e^V e^{-V}},  
 \label{dscq:sf}
\eea
where $(\Sgm,V)=2ig(\Sgm^I,V^I)t_I$, 
and 
\be
 \tl{N} = N-\frac{i}{2}\brkt{Y-\bar{Y}}. 
\ee

Here we modify some quantities by adding terms involving $U^4$. 
We redefine $\cU$ in (\ref{def:cU}) as 
\be
 \cU(\Phi) = \brkt{1+iU^\mu\der_\mu+iU^4\der_y}\Phi, \;\;\;\;\;
 \cU(\bar{\Phi}) = \brkt{1-iU^\mu\der_\mu-iU^4\der_y}\bar{\Phi}, 
 \label{modify:cU}
\ee
for a chiral superfield~$\Phi$. 
In the following, we assume the Abelian gauge group, for simplicity. 
Then we modify $X^I$ in (\ref{def:cW^I}) as 
\be
 X^I \equiv \brkt{1+\frac{1}{4}U^\mu\bar{\sgm}_\mu^{\dalp\alp}
 \sbk{D_\alp,\bar{D}_{\dalp}}}V^I
 -iU^4\brkt{\Sgm^I-\bar{\Sgm}^I}, 
\ee
so that the gauge transformation law~$\dgt X^I=\Lmd^I+\bar{\Lmd}^I$ 
is maintained. 
This modification does not change the $\dscp$-transformation law of it, 
and under the $\dscq$-transformation, 
\be
 \dscq X^I = \frac{1}{2}\brkt{Y+\bar{Y}}\der_y X^I
 +\frac{1}{2}\brkt{Y-\bar{Y}}\brkt{\Sgm^I-\bar{\Sgm}^I}. 
\ee 
With this definition of $X^I$, $\dscq\cW_\alp^I$ does not depend on $N$. 
We also redefine the other field strength superfield~$\cV^I$ 
and $\cC_\mu$ in (\ref{cL_cC}) 
so that their $\dscq$-transformation are independent of $N$, 
\bea
%
 \cV^I  \defa \brkt{1-\frac{1}{4}U^\mu\bar{\sgm}_\mu^{\dalp\alp}
 \sbk{D_\alp,\bar{D}_{\dalp}}}
 \brkt{-\hat{\der}_yX^I+\Sgm^I+\bar{\Sgm}^I} 
 -\brc{\frac{i}{2}D^\alp\brkt{U^4\cW_\alp^I}+\hc} \nonumber\\
 \eql -\brkt{\hat{\der}_y+\frac{1}{4}\der_y U^\mu\bar{\sgm}^{\dalp\alp}_\mu
 \sbk{D_\alp,\bar{D}_{\dalp}}}V^I+\cU(\Sgm^I)+\cU(\bar{\Sgm}^I)
 +i\der_y U^4\brkt{\Sgm^I-\bar{\Sgm}^I} \nonumber\\
 &&-\brkt{\frac{i}{2}D^\alp U^4\cW_\alp^I+\hc}, \nonumber\\
 \cC_\mu \defa \der_y U_\mu+\frac{1}{2}\bar{\sgm}_\mu^{\dalp\alp}
 \brkt{\bar{D}_{\dalp}\Psi_\alp-D_\alp\bar{\Psi}_{\dalp}}+\der_\mu U^4. 
\eea
The redefined field strength superfields transform as 
\bea
 \dscq\cW_\alp^I \eql \frac{1}{4}\bar{D}^2D_\alp\brc{
 \frac{1}{2}\brkt{Y+\bar{Y}}\cV^I}, \nonumber\\
 \dscq\cV^I \eql \der_y\brc{\frac{1}{2}\brkt{Y+\bar{Y}}\cV^I}
 +\brkt{\frac{1}{4}D^\alp Y\cW_\alp^I+\hc}, \nonumber\\
 \dscq\cC_\mu \eql \frac{i}{2}\der_\mu\brkt{Y-\bar{Y}}.   \label{dscq:cWV}
\eea

As shown in Appendix~\ref{Action_invariance}, the $\dscq$-variations of 
the $d^4\tht$-integral and the $d^2\tht$-integral parts of the Lagrangian 
are cancelled at the zeroth order in the gravitational superfields, 
\be
 \dscq\sbk{2\int\dr^4\tht\;\Omg+\brc{\int\dr^2\tht\;W+\hc}} = 0, 
 \label{cancellation}
\ee
where $\Omg\equiv\Omg^{\rm hyper}+\Omg^{\rm vector}$ 
and $W\equiv W^{\rm hyper}+W^{\rm CS}$. 
We have dropped total derivatives. 

In order to determine a constant~$b$ in (\ref{cL_cC}), we now take into account 
linear order contributions to 
the matter-independent part of $\dscq\cL$ in the gravitational superfields, 
as we did in the 4D case discussed in Sec.~\ref{L_kin:U}. 
Here note that $U^4$ can be completely gauged away 
by the $N$-transformation. 
We will take such a gauge in the following. 
Next, in order to pick up the matter-independent part of the action, 
we expand chiral superfields~$\Phi_{\rm odd}$, $\Phi_{\rm even}$ and $\Sgm^I$ 
around the backgrounds as 
\be
 \Phi_{\rm odd} = \vev{\Phi_{\rm odd}}+\tl{\Phi}_{\rm odd}, \;\;\;\;\;
 \Phi_{\rm even} = \vev{\Phi_{\rm even}}+\tl{\Phi}_{\rm even}, \;\;\;\;\;
 \Sgm^I = \vev{\Sgm^I}+\tl{\Sgm}^I.  \label{bg:chiral}
\ee
where the backgrounds~$\vev{\Phi_{\rm odd}}$, 
$\vev{\Phi_{\rm even}}$, and $\vev{\Sgm^I}$ 
are functions of only $y$, and satisfy 
\bea
 &&\vev{\Phi_{\rm odd}}^\dagger\tl{d}\vev{\Phi_{\rm odd}}
 +\vev{\Phi_{\rm even}}^\dagger\tl{d}\vev{\Phi_{\rm even}}  
 = 1, \nonumber\\
 &&C_{IJK}(2\Re\vev{\Sgm^I})(2\Re\vev{\Sgm^J})
 (2\Re\vev{\Sgm^K}) = 1.  \label{bg:D-gauge}
\eea
These conditions come from the $\bdm{D}$ gauge-fixing, 
which we will explain in Sec.~\ref{unphysical_mode}. 
Then, we obtain 
\bea
 \dscq\Omg^{\rm hyper} \eql -\frac{1}{2}\der_y\brkt{Y+\bar{Y}}+\cdots, 
 \nonumber\\
 \dscq\Omg^{\rm vector} \eql -\frac{1}{4}\der_y\brkt{Y+\bar{Y}}
 +\frac{3i}{4}\der_y\brkt{Y-\bar{Y}}\der_\mu U^\mu+\cdots, \nonumber\\
 \dscq W^{\rm hyper} \eql \frac{i}{8}\bar{D}^2
 \brkt{\der^\mu\bar{Y}\bar{\sgm}_\mu^{\dalp\alp}\bar{D}_{\dalp}\Psi_\alp}
 +\cdots, \nonumber\\
 \dscq W^{\rm CS} \eql 0+\cdots,  \label{dscq:Omg-W2}
\eea
where the ellipses denote matter-dependent terms and 
higher-order terms in the gravitational superfields. 
We have used (\ref{bg:D-gauge}). 
The $\Psi$-dependent term appears 
since $\der_y$ in $\dscq W^{\rm hyper}$ in (\ref{dscq:Omg-W}) 
should be understood as $\hat{\der}_y$ when we take into account 
the linear order contributions in the gravitational superfields. 
Thus the $\dscq$-variation of the 5D Lagrangian with (\ref{cL_cC}), 
\be
 \cL = \int\dr^4\tht\;\brc{E_2+2b\cC^\mu\cC_\mu
 +2\brkt{1+\frac{1}{3}E_1}\Omg}
 +\brc{\int\dr^2\tht\;W+\hc}, 
\ee
is calculated as 
\bea
 \dscq\cL \eql \int\dr^4\tht\;\left[4b\cC^\mu\dscq\cC_\mu
 -\frac{1}{2}\der_y\brkt{Y+\bar{Y}}E_1  \right. \nonumber\\
 &&\left. \hspace{15mm}
 +\frac{i}{2}\der^\mu\brkt{Y-\bar{Y}}\brc{3\der_y U_\mu 
 +\bar{\sgm}_\mu^{\dalp\alp}
 \brkt{\bar{D}_{\dalp}\Psi_\alp-D_\alp\bar{\Psi}_{\dalp}}}\right]+\cdots 
 \nonumber\\
 \eql \int\dr^4\tht\;
 \sbk{\frac{i}{2}\der^\mu\brkt{Y-\bar{Y}}
 \brkt{4b\cC_\mu+2\cC_\mu}}+\cdots. 
\eea
where the ellipsis denote matter-dependent terms 
and higher-order terms in the gravitational superfields. 
We have dropped total derivatives. 
Therefore, the $\dscq$-invariance of the action requires that 
$b=-\frac{1}{2}$. 

\ignore{
Let us consider a matter-independent part of $\dscq\Omg$, 
including linear terms in $E_{\rm grav}$.  
\be
 \dscq\Omg = \frac{1}{2}\der_y\brkt{Y+\bar{Y}}\Omg_{\rm c}
 +\Dlt(\dscq\Omg)+\cdots, 
 \label{dscq:Omg}
\ee
where $\Dlt(\dscq\Omg)$ denotes linear terms in $E_{\rm grav}$, 
and the ellipsis denotes matter-dependent terms.  
Note that 
$\cO(E_{\rm grav}^2)$-corrections to the components of the matter superfields, 
which have been neglected, contribute to $\Dlt(\dscq\Omg)$. 
Including such corrections, the components of $\Omg$ are expressed 
as (\ref{comp:Omg}), but now $\Dlt\phi^\Omg$ 
($\phi^\Omg=C^\Omg,\zeta^\Omg_\alp,\cdots$) 
is quadratic in the 5D gravitational fields. 
The components of $\tl{V}_E$ do not appear in $\Dlt\phi^\Omg$ 
because their dependence of the action is already specified 
as in (\ref{V_E-dependence}). 
Thus we classify the rest of $E_{\rm grav}$ according to the $Z_2$-parity, 
and denote the $Z_2$-even and odd fields as $E_{\rm even}$ and $E_{\rm odd}$, 
respectively. 
Namely, $E_{\rm even}$ and $E_{\rm odd}$ are the components of 
$U^\mu$ and $(\Psi_\alp,U^4)$. 
Similarly, we collectively denote the components of $L$ and $(Y,N)$ 
as $\Xi_{\rm even}$ and $\Xi_{\rm odd}$, respectively. 
Then, the transformation laws of $E_{\rm even}$ and $E_{\rm odd}$ are given by 
\bea
 \dscp E_{\rm even} \eql \cO(\Xi_{\rm even}), \;\;\;\;\;
 \dscp E_{\rm odd} = \cO(\der_y\Xi_{\rm even}), \nonumber\\
 \dscq E_{\rm even} \eql 0, \;\;\;\;\;
 \dscq E_{\rm odd} = \cO(\Xi_{\rm odd}),  \label{dsc:Es}
\eea
up to the zeroth order in the fields. 
Now $\Dlt\phi^\Omg$ is devided into three parts.  
\be
 \Dlt\phi^\Omg = 
 \Dlt_1\phi^\Omg+\Dlt_2\phi^\Omg+\Dlt_3\phi^\Omg, 
\ee
where $\Dlt_1\phi^\Omg=\cO(E_{\rm even}^2)$, 
$\Dlt_2\phi^\Omg=\cO(E_{\rm even}E_{\rm odd})$, 
and $\Dlt_3\phi^\Omg=\cO(E_{\rm odd}^2)$. 
Similarly to the 4D case, $\Dlt\phi^\Omg$ is determined 
so that the matter-independent part of 
$\dscp\Omg$ does not have linear terms in $E_{\rm even}$. 
As we mentioned in Sec.~\ref{L_kin:U}, 
$\Dlt_1\phi^\Omg$ already satisfies this condition. 
Therefore, $\Dlt_2\phi^\Omg$ must vanish since 
$\dscp\Dlt_2\phi^\Omg$ also contains linear terms in $E_{\rm even}$. 
As a result, we find that $\Dlt(\dscq\Omg)$ does not contain 
linear term in $E_{\rm even}$. 
}

\subsection{Identification of components in $\bdm{\Psi_\alp}$ and $\bdm{U^4}$} 
\label{ID:Z2odd}
In this subsection, we identify the components in 
the $Z_2$-odd gravitational superfields~$\Psi_\alp$ and $U^4$ 
with the component fields in Ref.~\cite{Kugo:2002js}. 
By comparing (\ref{dscq:sf}) with the superconformal transformations 
in Ref.~\cite{Kugo:2002js}, we identify components of $Y$ and $N$ as 
\bea
 Y \eql \xi^4+4\tht\ep_\alp^-+i\tht^2\brkt{\vth_V^1+i\vth_V^2}, 
 \nonumber\\
 N \eql \brkt{\tht\sgm^\mu\bar{\tht}}\brkt{\lmd_{\mu 4}+\der_\mu\xi^4}
 +\tht^2\bar{\tht}\brkt{\bar{\eta}^+-\frac{1}{4}\bar{\sgm}^\mu\der_\mu\ep^-}
 +\bar{\tht}^2\tht\brkt{\eta^++\frac{1}{4}\sgm^\mu\der_\mu\bar{\ep}^-} \nonumber\\
 &&+\cdots, 
\eea
where we take the chiral coordinate~$y^\mu$ for $Y$. 
The parameters~$\xi^4$, $\lmd_{\mu 4}$, and $\vth_V^r$ ($r=1,2$) 
are the transformation parameters 
for $\bdm{P^4}$, $\bdm{M_{\mu 4}}$, and $\suU/U(1)_A$. 
The spinors~$\ep_\alp^-$ and $\eta^+_\alp$ are 
the $Z_2$-odd components of the transformation parameters 
for $\bdm{Q}$ and $\bdm{S}$.  

Therefore, each component of $\Psi_\alp$ and $U^4$ are identified as 
\bea
 \left.\bar{D}_{\dalp}\Psi_\alp\right|_0 
 \eql -\frac{i}{2}\sgm^\mu_{\alp\dalp}e_{y\underline{\mu}}, \;\;\;\;\;
 \left.D^\alp\bar{D}_{\dalp}\Psi_\alp\right|_0 
 = \frac{3i}{2}\brkt{\psi_\mu^-\sgm^\mu}_{\dalp}+4\bar{\psi}_{y\dalp}^+, 
 \nonumber\\
 \left.D^2\bar{D}_{\dalp}\Psi_\alp\right|_0 
 \eql -4\sgm^\mu_{\alp\dalp}\brkt{V_\mu^1+iV_\mu^2}, \;\;\;\;\;
 \cdots, 
\eea
and 
\be
 U^4 = \brkt{\tht\sgm^\mu\bar{\tht}}e_\mu^{\;\;4}
 +\frac{1}{4}\bar{\tht}^2\brkt{\tht\sgm^\mu\bar{\psi}_\mu^-}
 -\frac{1}{4}\tht^2\brkt{\bar{\tht}\bar{\sgm}^\mu\psi_\mu^-}
 +\cdots, 
\ee
Notice that $\Psi_\alp$ appears in the action only through 
$\bar{D}_{\dalp}\Psi_\alp$ and its derivatives. 
Thus $\Psi_\alp|_0$ and $D_\alp\Psi_\bt|_0$ are irrelevant to the physics. 

There is one comment on the component identification for the matter superfields. 
As mentioned in Sec.~\ref{decomp:sf}, 
each component of the $\cN=1$ superfields listed in Appendix~\ref{comp:superfields} 
may be corrected by terms involving the $Z_2$-odd fields. 
In fact, we need such correction terms in order to reproduce 
the correct superconformal transformations in Ref.~\cite{Kugo:2002js}. 
For example, $\zeta_\alp^{a\pm}$ in (\ref{expr:Phi^oa}) must be modified as 
\bea
 \zeta_\alp^{a-} \toa \zeta_\alp^{a-}
 -\frac{7}{8}\brkt{\sgm^\mu\bar{\psi}_\mu^-}_\alp\bar{\cA}_2^{2a}, 
 \nonumber\\
 \zeta_\alp^{a+} \toa \zeta_\alp^{a+}
 +\frac{7}{8}\brkt{\sgm^\mu\bar{\psi}_\mu^-}_\alp\bar{\cA}_2^{2a-1}. 
\eea

\subsection{Elimination of unphysical modes}  \label{unphysical_mode}
Since our action is based on the superconformal formulation, 
there are unphysical degrees of freedom to eliminate. 

As pointed out in Ref.~\cite{Correia:2006pj}, 
$V_E$ does not have a kinetic term and can be integrated out.\footnote{
This does not mean that $e_y^{\;\;4}$ is an auxiliary field. 
It is also contained in $\Sgm^I$, which have their own kinetic terms. } 
From (\ref{V_E-dependence}), $V_E$ is expressed as 
\be
 V_E = \brkt{\frac{\Omg^{\rm v}}{\Omg^{\rm h}}}^{1/3}, 
 \label{expr:V_E}
\ee
where
\bea
 \Omg^{\rm h} \defa -\Omg^{\rm hyper}|_{V_E=1}
 = \cU(\Phi^\dagger_{\rm odd})\tl{d}\brkt{e^V}^t\cU(\Phi_{\rm odd})
 +\cU(\Phi^\dagger_{\rm even})\tl{d}e^{-V}\cU(\Phi_{\rm even}), \nonumber\\
 \Omg^{\rm v} \defa -2\Omg^{\rm vector}|_{V_E=1} 
 = C_{IJK}\cV^I\cV^J\cV^K 
 = -\frac{c^3}{24g^3}\tr\brkt{\cV^3}. 
\eea
After integrating it out, $\Omg$ in the action becomes 
\be
 \Omg = -\frac{3}{2}\brkt{\Omg^{\rm h}}^{2/3}\brkt{\Omg^{\rm v}}^{1/3}.  
 \label{final:Omg}
\ee

In order to obtain the Poincar\'{e} SUGRA, we have to impose
the superconformal gauge fixing conditions to  eliminate the extra symmetries. 
The following conditions imposed on the hypermultiplet sector fix 
those symmetries. 
\bea
 d_{\oa}^{\;\;\ob}\bar{\phi}_{\ob}\phi^{\oa} 
 \eql \phi_{\rm odd}^\dagger\tl{d}\phi_{\rm odd}
 +\phi_{\rm even}^\dagger\tl{d}\phi_{\rm even} = 1, \nonumber\\
 d_{\oa}^{\;\;\ob}\bar{\phi}_{\ob}\chi^{\oa} 
 \eql \phi_{\rm odd}^\dagger\tl{d}\chi_{\rm odd}
 +\phi_{\rm even}^\dagger\tl{d}\chi_{\rm even} = 0, \nonumber\\
 \phi^{\oa}d_{\oa}^{\;\;\ob}\rho_{\ob\oc}\chi^{\oc} 
 \eql \phi_{\rm odd}\tl{d}\chi_{\rm even}-\phi_{\rm even}\tl{d}\chi_{\rm odd} 
 = 0, \nonumber\\
 \phi^1 \eql 0, \;\;\;\;\;
 \arg(\phi^2) = 0,  \label{scGF_cond:5D}
\eea
in the unit of the 5D Planck mass~$M_5=1$. 
Here $\sbk{\phi^{\oa},\chi^{\oa},\cdots}$ is a chiral multiplet 
corresponding to a chiral superfield~$\Phi^{\oa}$.  
The first condition fixes the $\bdm{D}$ gauge, 
the second and the third fix the $\bdm{S}$ gauge, 
and the fourth and the fifth fix the $\suU$ gauge. 
The $\bdm{K}$ gauge is already fixed in our formalism. 
In the single compensator case~$n_C=1$, for example, 
They are rewritten as 
\bea
 \phi^1 \eql 0, \;\;\;\;\;
 \phi^2 = \brc{1+\sum_{\oa=3}^{2(n_H+1)}\abs{\phi^{\oa}}^2}^{1/2}, \nonumber\\
 \chi^1_\alp \eql -\brkt{\phi^2}^{-1}\sum_{\oa,\ob=3}^{2(n_H+1)}
 \rho_{\oa\ob}\phi^{\oa}\chi^{\ob}_\alp, \;\;\;\;\;
 \chi^2_\alp = \brkt{\bar{\phi}^2}^{-1}
 \sum_{\oa=3}^{2(n_H+1)}\bar{\phi}^{\oa}\chi^{\oa}.  
\eea
The components of the compensator hypermultiplet are now expressed 
in terms of the physical fields. 

Instead of the first two conditions in (\ref{scGF_cond:5D}), 
we can also take the following gauge fixing conditions. 
\be
 \Omg^{\rm h}|_0 = 1, \;\;\;\;\;
 D_\alp\Omg^{\rm h}|_0 = 0.  \label{scGF_cond:5D:2}
\ee
Notice that these are different from (\ref{scGF_cond:5D}) 
due to the existence of $\cE$ in (\ref{sf:chiral}). 
These conditions can be rewritten as 
\be
 \left.\frac{\Omg^{\rm v}}{V_E^3}\right|_0 = 1, \;\;\;\;\;
 \left.D_\alp\brkt{\frac{\Omg^{\rm v}}{V_E^3}}\right|_0 = 0, 
 \label{scGF_cond:5D:3}
\ee
or 
\be
 \Omg|_0 = -\frac{3}{2}V_E|_0, \;\;\;\;\;
 D_\alp\Omg|_0 = -\frac{3}{2}D_\alp V_E|_0. 
\ee
We have used (\ref{expr:V_E}). 

The expression~(\ref{final:Omg}) is useful to compare our superfield Lagrangian 
to that of Ref.~\cite{Linch:2002wg}. 
For simplicity, we consider the single compensator case. 
In this case, $\Phi^2$ plays a similar role to the chiral compensator 
multiplet in 4D SUGRA. 
To emphasize this point, we redefine $\Phi^{\oa}$ as 
$\Phi_C\equiv\brkt{\Phi^2}^{2/3}$ and $\hat{\Phi}^{\oa}\equiv\Phi^{\oa}/\Phi^2$ 
($\oa\neq 2$) so that their Weyl weights are one and zero, respectively. 
In a case that the background values of physical matter fields 
are zero or negligible compared with $M_5$, (\ref{bg:chiral}) is rewritten as 
\be
 \Phi_C = 1+\tl{\Phi}_C, \;\;\;\;\;
 \hat{\Phi}^{\oa} = 0+\cdots, \;\;\;\;\;
 \Sgm^I = \vev{\Sgm^I}+\tl{\Sgm}^I, 
\ee
where $\oa\neq 2$, 
and the ellipsis denotes terms involving the physical matter fields. 
Then, (\ref{final:Omg}) is rewritten as 
\bea
 \Omg \eql -\frac{3}{2}\abs{\cU(\Phi_C)}^2\brkt{1+\cdots}^{2/3}
 \brc{1+8C_{IJK}(\Re\vev{\Sgm^I})(\Re\vev{\Sgm^J})
 \Re\cU(\tl{\Sgm}^K)+\cdots} \nonumber\\
 \eql -\frac{3}{2}\abs{1+\tl{\Phi}_C+iU^\mu\der_\mu\tl{\Phi}_C}^2
 \brc{1+T+\bar{T}+iU^\mu\der_\mu (T-\bar{T})+\cdots}, 
\eea
where $T\equiv 4C_{IJK}(\Re\vev{\Sgm^I})(\Re\vev{\Sgm^J})\tl{\Sgm}^K$ is 
(the fluctuation of) the 5D radion superfield~\cite{Abe:2011rg}. 
Then the 5D Lagrangian is 
\bea
 \cL \eql \int\dr^4\tht\;\brc{E_2-\cC^\mu\cC_\mu
 +2\brkt{1+\frac{1}{3}E_1}\Omg}+\cdots \nonumber\\
 \eql \int\dr^4\tht\;\left[E_2-\cC^\mu\cC_\mu
 -3\abs{\tl{\Phi}_C}^2-3iU^\mu\der_\mu\brkt{\tl{\Phi}_C+T+\hc} \right. \nonumber\\
 &&\left. \hspace{15mm}
 -E_1\brkt{\tl{\Phi}_C+T+\hc}-3\brkt{\bar{\tl{\Phi}}_CT+\hc}\right]+\cdots 
 \nonumber\\
 \eql \int\dr^4\tht\;\left[E_2-3\abs{\tl{\Phi}_C}^2
 +2i\der_\mu U^\mu\brkt{\tl{\Phi}_C-\bar{\tl{\Phi}}_C} \right. \nonumber\\
 &&\left. \hspace{15mm}
 -\cC^\mu\cC_\mu
 +\brc{\brkt{-3\bar{\tl{\Phi}}_C+2i\der_\mu U^\mu}T+\hc}\right]+\cdots. 
\eea
We have dropped total derivatives. 
Recalling that the chiral compensator~$\Sgm^{\rm cp}$ in Ref.~\cite{Linch:2002wg} 
is identified with $\Sgm^{\rm cp}=3\tl{\Phi}_C$, 
this agrees with (3.19) and (3.20) of Ref.~\cite{Linch:2002wg}, 
except for the second line in (3.20) of Ref.~\cite{Linch:2002wg}. 
There is no counterpart of the latter in our formulation  
because we do not need the prepotential for $\Sgm^{\rm cp}$ 
to construct an invariant action. 
In Ref.~\cite{Linch:2002wg}, 
the role of the second line in (3.20) 
is to cancel the $\dscq$-variation of the first and the third lines. 
In our formulation, such variation is cancelled by 
$\dscq\Omg$ and $\dscq W$ in (\ref{dscq:Omg-W2}).

\subsection{Case of warped geometry} \label{warped_case}
We have assumed the 5D flat spacetime~(\ref{flat_metric}) 
as a background geometry. 
Here we extend the results obtained so far 
to a warped geometry whose metric is given by 
\be
 ds^2 = e^{2\sgm(y)}\eta_{\mu\nu}dx^\mu dx^\nu-dy^2.  \label{warped_metric}
\ee
This is the most generic metric that has the 4D Poincar\'e symmetry. 
The warp factor~$\sgm(y)$ is determined by solving the equations of motion. 
Especially, for a case of a supersymmetric background, it is obtained 
as a solution to the BPS equations. 
(See, for example, Ref.~\cite{Abe:2006eg}.)
A nontrivial warp factor is obtained in a case that the compensator hypermultiplets 
are charged under some of the gauge groups, 
\ie, the gauged SUGRA~\cite{Lukas:1998yy,Falkowski:2000yq,Abe:2007zv}. 
The warp factor can be easily taken into account in the following way. 
First we rescale each component field by $e^{-w\sgm}$, 
where $w$ is its Weyl weight~\cite{Paccetti:2004ri}. 
Second we change the coordinate~$y$ as 
\be
 d\hat{y} = e^{-\sgm(y)}dy.  \label{def:haty}
\ee
Then, the above background metric becomes the flat one, 
\be
 ds^2 = \eta_{\mu\nu}dx^\mu dx^\nu-d\hat{y}^2. 
\ee
This means that we can obtain the superfield action 
in the warped spacetime from that in the flat spacetime 
by going back to the $y$-coordinate and 
rescaling each component field by $e^{w\sgm}$. 
For example, $e_y^{\;\;4}$ transforms under such transformations as 
\be
 e_y^{\;\;4} \to e^{\sgm}e_y^{\;\;4} \to e_y^{\;\;4}, 
\ee
where the first transformation is by $\hat{y}\to y$ 
and the second one is the Weyl rescaling. 
Thus the background value of $V_E$ remains to be one 
after these transformations. 
It is convenient to further rescale the chiral superfields~$\Phi^{\oa}$ 
and $\Sgm^I$ as 
\be
 \Phi^{\oa} \to e^{\frac{3}{2}\sgm}\Phi^{\oa}, \;\;\;\;\;
 \Sgm^I \to e^{\sgm}\Sgm^I, 
\ee
so that the lowest components of the superfields do not contain the warp factor. 
Then the $\bdm{D}$ gauge fixing condition in (\ref{scGF_cond:5D:2}) 
or (\ref{scGF_cond:5D:3}) is unchanged. 
The covariant derivative~$\hat{\der}_y$ transforms as 
$\hat{\der}_y\to e^\sgm\hat{\der}_y$, where it is now defined as 
\be
 \hat{\der}_y \equiv \der_y
 -e^{-\sgm}\brkt{\frac{1}{4}\bar{D}^2\Psi^\alp D_\alp
 +\frac{1}{2}\bar{D}^{\dalp}\Psi^\alp\bar{D}_{\dalp}D_\alp
 +\frac{w+n}{24}\bar{D}^2D^\alp\Psi_\alp+\hc}.  \label{def:der_haty}
\ee
Notice that 
\be
 \Omg \equiv -V_E\Omg^{\rm h}-\frac{1}{2}V_E^{-2}\Omg^{\rm v} 
\ee
transforms as $\Omg\to e^{3\sgm}\Omg$. 
This indicates that the coefficient of the kinetic terms 
for the SUGRA superfields is also rescaled by $e^{3\sgm}$. 
As a result, the 5D Lagrangian is 
\bea
 \cL \eql \int\dr^4\tht\;e^{2\sgm}\brc{E_2-\cC^\mu\cC_\mu
 +2\brkt{1+\frac{1}{3}E_1}\Omg} \nonumber\\
 &&+\sbk{\int\dr^2\tht\;\brkt{e^{3\sgm}W^{\rm hyper}+W^{\rm CS}}+\hc}, 
 \label{cL:warped}
\eea
where
\be
 \cC_\mu \equiv e^\sgm\der_y U_\mu+\frac{1}{2}\bar{\sgm}_\mu^{\dalp\alp}
 \brkt{\bar{D}_{\dalp}\Psi_\alp-D_\alp\bar{\Psi}_{\dalp}}, 
\ee
and the definitions of the other quantities are unchanged, 
except for $\hat{\der}_y$ defined in (\ref{def:der_haty}). 
We have taken into account the warp factor coming from 
the integral measure~(\ref{def:haty}).

\ignore{
the rescaled f\"unfbein becomes 
\bea
 e_\mu^{\;\;\underline{\nu}} \eql \dlt_\mu^{\;\;\nu}+e^{-\sgm}\tl{e}_\mu^{\;\;\nu}, 
 \;\;\;\;\;
 e_\mu^{\;\;4} = e^{-\sgm}\tl{e}_\mu^{\;\;4}, \nonumber\\
 e_y^{\;\;\underline{\nu}} \eql e^{-\sgm}\tl{e}_y^{\;\;\nu}, \;\;\;\;\;
 e_y^{\;\;4} = e^{-\sgm}+e^{-\sgm}\tl{e}_y^{\;\;4}. 
\eea
Since the 4D part of the background metric becomes the 4D flat one  
after this rescaling, the results in the previous subsections remain valid 
by regarding component fields of each superfield as the rescaled one. 
The background value of $V_E$ is now $e^{-\sgm}$. 
So we define $\tl{V}_E$ as $V_E\equiv e^{-\sgm}(1+\tl{V}_E)$. 
It is also convenient to further rescale $\Phi^{\oa}$ as 
$\Phi^{\oa}\to e^{\frac{3}{2}\sgm}\Phi^{\oa}$ 
so that the $\bdm{D}$-gauge fixing condition is still expressed 
as (\ref{scGF_cond:5D}) or (\ref{scGF_cond:5D:2}). 
This leads to $\Omg^{\rm h}\to e^{3\sgm}\Omg^{\rm h}$, $\Omg\to e^{2\sgm}\Omg$, 
$W^{\rm hyper}\to e^{3\sgm}W^{\rm hyper}$, and  
the resultant Lagrangian is written as 
\bea
 \cL \eql \int\dr^4\tht\;
 \brc{E_2-\cC^\mu\cC_\mu
 +2e^{2\sgm}\brkt{1+\frac{1}{3}E_1}\Omg} \nonumber\\
 &&+\sbk{\int\dr^2\tht\;\brkt{e^{3\sgm}W^{\rm hyper}+W^{\rm CS}}+\hc}, 
 \label{cL:warped}
\eea
where $W^{\rm hyper}$, $W^{\rm CS}$ and $\Omg$ are 
defined in (\ref{cL^hyper:NA}), (\ref{expr:W^CS:NA}) 
and (\ref{final:Omg}), respectively.  
}

\section{Summary} \label{summary}
We have completed an $\cN=1$ superfield action for the generic system 
of vector multiplets and hypermultiplets coupled to 5D SUGRA,  
based on the superconformal formulation~\cite{Kugo:2000af}-\cite{Kugo:2002js}. 
Especially we specified couplings of the gravitational superfields 
to the matter superfields, up to linear order in the former. 
The gravitational superfields consist of 
four superfields~$(U^\mu, \tl{V}_E, \Psi_\alp, U^4)$, 
which correspond to the fluctuation modes 
around the background metric~(\ref{flat_metric}) or (\ref{warped_metric}). 
The dependence of the action on these superfields is 
uniquely determined by 
the invariance under the 5D superconformal transformations~$\dscp$, $\dscq$, 
and the (super)gauge transformation~$\dgt$, 
which are expressed in the $\cN=1$ superfield description. 
Among them, 
$\cN=1$ part of the 5D superconformal transformation~$\dscp$ mainly restricts 
the form of the action. 
The others are used to fix the coefficients 
$a$ in (\ref{expr:cZ^IJ}) and $b$ in (\ref{cL_cC}). 
Since $\Psi_\alp$ and $U^4$ can be gauged away 
by the $\dscp$- and $\dscq$-transformations, respectively,  
and $\tl{V}_E$ can be integrated out as in (\ref{expr:V_E}), 
only $U^\mu$ remains in the unitary gauge.  
Notice that the matter multiplets need the help of the gravitational fields 
to form $\cN=1$ superfields~$\Phi^{\oa}$, $V^I$ and $\Sgm^I$, 
and thus they contain the gravitational fields. 
This means that the kinetic terms for the gravitational superfields, 
\ie, the first two terms in the first line of (\ref{cL:warped}), 
do not reproduce the Einstein-Hilbert term by themselves. 

This work can be understood as an extension of 
the 5D linearized SUGRA~\cite{Linch:2002wg} 
to a case that the matter superfields also propagate in the bulk. 
The 5D linearized SUGRA is useful to calculate quantum effects, 
keeping the $\cN=1$ superfield structure~\cite{Buchbinder:2003qu,Gregoire:2004nn}. 
Our result makes it possible to 
perform such calculations in more generic 5D SUGRA. 
For example, 
one-loop corrections to the effective 4D K\"ahler potential 
have to be calculated when we discuss the stabilization of 
the size of the extra dimension 
by the Casimir energy~\cite{Fabinger:2000jd,Garriga:2000jb} 
in the context of 5D SUGRA. 
An investigation of such moduli stabilization 
by means of the superfield action obtained in this paper 
is one of our future projects. 

Another direction to proceed is an extension of our formalism 
to higher-dimensional SUGRA. 
Although such theories do not have a full superconformal description, 
the $\cN=1$ superfield description is possible. 
(See Ref.~\cite{ArkaniHamed:2001tb,Abe:2012ya}, in the global SUSY case.)
Such superfield description provides a powerful tool 
to discuss SUSY brane-world models. 


\subsection*{Acknowledgements}
This work was supported in part by 
Grant-in-Aid for Young Scientists (B) No.22740187 
from Japan Society for the Promotion of Science.

\appendix

\section{Transformation law of field strength superfield} \label{Trf:fss}
Here we show that the field strength superfields defined by (\ref{def:cW}) 
and (\ref{def:cW:NA}) 
correctly transform as (superconformal) chiral multiplets with 
the Weyl weight $3/2$. 

First, consider the Abelian case. 
From (\ref{dsc_trf}) and (\ref{def:cW}), we find that 
\be
 \dsc\cW_\alp = \brkt{
 -\frac{1}{4}\bar{D}^2L^\bt D_\alp
 +\frac{1}{4}\bar{D}^2L_\alp D^\bt
 -i\sgm^\mu_{\alp\dbt}\bar{D}^{\dbt}L^\bt\der_\mu
 -i\bar{\sgm}^{\mu\dbt\bt}\bar{D}_{\dbt}L_\alp\der_\mu
 +\frac{1}{4}\bar{D}^2D_\alp L^\bt}\cW_\bt,  \label{dsc:cW1}
\ee
and $\bar{L}_{\dalp}$-dependent terms are certainly cancelled. 
Here we have used that 
$D^\bt\cW_\bt=\bar{D}_{\dbt}\bar{\cW}^{\dbt}$ and 
$\bar{D}^2\bar{\cW}^{\dbt} = -4i\bar{\sgm}^{\mu\dbt\bt}\der_\mu\cW_\bt$
at the zeroth order in $U^\mu$. 

In general, a matrix~$T_\alp^{\;\;\bt}$ can be expanded as 
\be
 T_\alp^{\;\;\bt} = \frac{1}{2}T_\gm^{\;\;\gm}\dlt_\alp^{\;\;\bt}
 -\Re\brc{\brkt{\sgm_{\mu\nu}}_\gm^{\;\;\dlt}T_\dlt^{\;\;\gm}}
 \brkt{\sgm^{\mu\nu}}_\alp^{\;\;\bt}. 
\ee
Thus, 
\be
 T_\alp^{\;\;\bt}-T^\bt_{\;\;\alp} 
 = T_\alp^{\;\;\bt}-\ep^{\bt\gm}\ep_{\alp\dlt}T_\gm^{\;\;\dlt} 
 = T_\gm^{\;\;\gm}\dlt_\alp^{\;\;\bt}, 
\ee
since $\ep^{\bt\gm}\ep_{\alp\dlt}\brkt{\sgm^{\mu\nu}}_\gm^{\;\;\dlt}
=\brkt{\sgm^{\mu\nu}}_\alp^{\;\;\bt}$. 
Therefore, 
\bea
 &&-\frac{1}{4}\bar{D}^2L^\bt D_\alp+\frac{1}{4}\bar{D}^2L_\alp D^\bt 
 = -\frac{1}{4}\bar{D}^2L^\gm D_\gm \dlt_\alp^{\;\;\bt}, \nonumber\\
 &&-i\sgm^\mu_{\alp\dalp}\bar{D}^{\dalp}L^\bt
 -i\bar{\sgm}^{\mu\dbt\bt}\bar{D}_{\dbt}L_\alp 
 = -i\ep^{\bt\dlt}\bar{D}^{\dgm}\brkt{\sgm^\mu_{\alp\dgm}L_\dlt
 -\sgm^\mu_{\dlt\dgm}L_\alp} 
 = -i\sgm^\mu_{\gm\dgm}\bar{D}^{\dgm}L^\gm\dlt_\alp^{\;\;\bt}, \nonumber\\
 &&-\frac{1}{4}\bar{D}^2D_\alp L^\bt 
 = \frac{1}{8}\bar{D}^2D_\gm L^\gm\dlt_\alp^{\;\;\bt}
 -\Re\brc{\brkt{\sgm_{\mu\nu}}_\gm^{\;\;\dlt}\cdot
 \frac{1}{4}\bar{D}^2D_\dlt L^\gm}\brkt{\sgm^{\mu\nu}}_\alp^{\;\;\bt}. 
 \label{DL:relations}
\eea
By using these, (\ref{dsc:cW1}) is rewritten as 
\bea
 \dsc\cW_\alp \eql \brc{-\frac{1}{4}\bar{D}^2L^\bt D_\bt
 -i\sgm^\mu_{\bt\dbt}\bar{D}^{\dbt}L^\bt\der_\mu
 -\frac{1}{8}\bar{D}^2D^\bt L_\bt}\cW_\alp \nonumber\\
 &&+\frac{1}{4}\Re\brc{\brkt{\sgm_{\mu\nu}}_\bt^{\;\;\gm}
 \bar{D}^2D^\bt L_\gm}\brkt{\sgm^{\mu\nu}\cW}_\alp. \label{dsc:cW}
\eea
This is the correct superconformal transformation law of a chiral multiplet. 
The last term of the first line implies that $\cW_\alp$ have 
the Weyl weight~$3/2$, and the second line is a term proportional to
the Lorentz generator, which is absent in the case of scalar superfields. 
The chirality of the last term is not manifest, but it is ensured by 
the last equation in (\ref{DL:relations}). 
Namely, the $\bar{\tht}^{\dalp}$-dependent terms are cancelled 
by summing over the indices~$\mu$ and $\nu$. 

We can check the non-Abelian case in a similar way.  
From (\ref{dsc_trf}), we obtain 
\bea
 \dsc\sbk{e^V D_\alp e^{-V}}_{\bE} 
 \eql \brkt{-\frac{1}{4}\bar{D}^2L^\bt D_\bt-i\sgm^\mu_{\bt\dbt}
 \bar{D}^{\dbt}L^\bt\der_\mu}\brkt{e^VD_\alp e^{-V}} 
 -\frac{1}{4}\bar{D}^2D_\alp L^\bt e^V D_\alp e^{-V} \nonumber\\
 &&-\frac{1}{2}D_\alp\bar{D}^{\dbt}L^\bt\bar{D}_{\dbt}
 \brkt{e^V D_\bt e^{-V}}
 -i\sgm^\mu_{\alp\dbt}\der_\mu\bar{D}^{\dbt}L^\bt e^V D_\bt e^{-V}, 
\eea
which leads to 
\bea
 \dsc\cW_\alp = \brkt{-\frac{1}{4}\bar{D}^2L^\bt D_\bt
 -i\sgm^\mu_{\bt\dbt}\bar{D}^{\dbt}L^\bt\der_\mu}\cW_\alp
 +\frac{1}{4}\bar{D}^2D_\alp L^\bt\cW_\bt. 
\eea
With the last equation in (\ref{DL:relations}), 
this is the same transformation law as (\ref{dsc:cW}).

\ignore{
\section{Invariant action formulae in components}
In the superconformal formulation of 4D SUGRA, there are 
two superconformal-invariant action formulae, \ie, 
the $F$-term and $D$-term action formulae. 
For a chiral multiplet~$W=\sbk{\phi,\chi_\alp,F}$ with the Weyl weight~$w=3$, 
the $F$-term invariant action formula is given by 
\be
 S_F[W] = \int\dr^4x\;e\brkt{F-i\bar{\psi}_\mu\bar{\sgm}^\mu\chi+\hc
 +\cdots},  \label{F_formula}
\ee
where $e\equiv \det\brkt{e_\mu^{\;\;\underline{\nu}}}$, and the ellipsis 
denotes terms beyond the linear order in the SUGRA multiplet fields. 
For a real general multiplet~$\Omg=\sbk{C,\zeta_\alp,\cH,B_\mu,\lmd_\alp,D}$ 
with the Weyl weight~$w=2$, the $D$-term action formula is given by
\bea
 S_D[\Omg] \eql \int\dr^4x\;e\left[D-\bar{\psi}_\mu\bar{\sgm}^\mu\lmd
 +\psi_\mu\sgm^\mu\bar{\lmd}+\frac{4i}{3}\brkt{\zeta\sgm^{\mu\nu}\der_\mu\psi_\nu
 -\bar{\zeta}\bar{\sgm}^{\mu\nu}\der_\mu\bar{\psi}_\nu} \right. \nonumber\\
 &&\hspace{15mm}\left.
 +\frac{C}{3}\brc{R(\omg)+4\ep^{\mu\nu\rho\tau}
 \brkt{\psi_\mu\sgm_\tau\der_\nu\bar{\psi}_\rho
 -\bar{\psi}_\mu\bar{\sgm}_\tau\der_\nu\psi_\rho}}+\cdots\right], 
 \label{D_formula}
\eea
where $R(\omg)$ is the scalar curvature constructed from the spin connection. 
}

\section{Explicit forms of $\bdm{\cN=1}$ superfields in 5D SUGRA} 
\label{comp:superfields} 
Here we collect explicit expressions of $\cN=1$ superfields 
originating from 5D hyper and vector multiplets. 
We omit terms involving the $Z_2$-odd gravitational fields. 
Such terms need to be added to the following expressions 
in order to obtain the complete expressions, 
as mentioned in Sec.~\ref{ID:Z2odd}. 
We take the notations of Ref.~\cite{Kugo:2002js} 
for the component fields. 

A hypermultiplet~$[\cA_i^{\oa},\zeta^{\oa}_\alp,\cF^{\oa}_i]$ 
$(i=1,2;\oa=1,2,\cdots,2(n_C+n_H))$ is split into two chiral multiplets, 
\bea
 \Phi^{2a-1} \eql \brkt{1+\frac{1}{2}\cE}\brc{
 \cA_2^{2a-1}-2i\tht\zeta^{a-}+\tht^2 F^{2a-1}}, \nonumber\\
 \Phi^{2a} \eql \brkt{1+\frac{1}{2}\cE}\brc{
 \cA_2^{2a}-2i\tht\zeta^{a+}+\tht^2 F^{2a}},  \label{expr:Phi^oa}
\eea
where
\be
 F^{\oa} \equiv i\cF^{\oa}_1+\brkt{\cD_4\cA}_1^{\oa}
 +igM^I(t_I)^{\oa}_{\ob}\cA^{\ob}_1,  
\ee
and the definition of the covariant derivative~$(\cD_M\cA)_i^{\oa}$ 
is given in Ref.~\cite{Kugo:2002js}. 
The 2-component spinor notation of the hyperini, which are 
the symplectic-Majorana spinor, is defined from 
the 4-component notation in Ref.~\cite{Kugo:2002js} as 
\be
 \zeta^{2a-1} = \begin{pmatrix} \zeta^{a-} \\ -\bar{\zeta}^{a+} 
 \end{pmatrix}, \;\;\;\;\;
 \zeta^{2a} = \begin{pmatrix} \zeta^{a+} \\ \bar{\zeta}^{a-} 
 \end{pmatrix}. 
\ee

A 5D vector multiplet~$[M^I,W_M^I,\Omg^{Ii},Y^{Ir}]$ is decomposed into 
$\cN=1$ gauge and chiral superfields as 
\bea
 V^I \eql -\brkt{\tht\sgm^\mu\bar{\tht}}\brkt{e^{-1}}_\mu^{\;\;\nu}W_\nu^I
 +i\tht^2\bar{\tht}\brc{2\bar{\Omg}^{I+}
 -\brkt{\bar{\sgm}^\nu\sgm^\mu\bar{\psi}_\nu^+}W_\mu^I} \nonumber\\
 &&-i\bar{\tht}^2\tht\brc{2\Omg^{I+}-\brkt{\sgm^\nu\bar{\sgm}^\mu\psi_\nu^+}W_\mu^I}
 +\frac{1}{2}\tht^2\bar{\tht}^2D^I, \nonumber\\
 \Sgm^I \eql \frac{1}{2}\brkt{e_y^{\;\;4}M^I-iW_y^I}
 +\tht\brkt{2ie_y^{\;\;4}\Omg^{I-}+2\psi_y^- M^I} \nonumber\\
 &&-\tht^2\brc{e_y^{\;\;4}\brkt{Y^{I1}+iY^{I2}}
 +i\brkt{V_y^1+iV_y^2}M^I}, 
\eea
where 
\be
 D^I \equiv 2Y^{I3}-\brkt{\cD_4 M}^I
 +\brkt{-2\bar{\Omg}^{I+}\bar{\sgm}^\mu\psi_\mu^++\hc}
 +\brkt{2V^{3\mu}+2v^\mu_{\;\;4}
 -\frac{1}{2}\ep^{\mu\nu\rho\tau}\der_\nu\tl{e}_{\rho\tau}}W_\mu^I. 
\ee
The 2-component spinor notation of an $\suU$-Majorana 
spinor~$\psi^i=\Omg^{Ii},\psi^i_M$ 
is defined from the 4-component notation in Ref.~\cite{Kugo:2002js} as 
\be
 \psi^1 = \begin{pmatrix} \psi^+ \\ -\bar{\psi}^- \end{pmatrix}, \;\;\;\;\;
 \psi^2 = \begin{pmatrix} \psi^- \\ \bar{\psi}^+ \end{pmatrix}. 
 \label{rel:2-4spinor}
\ee

\section{Invariance of the action} \label{action_inv}
\subsection{Gauge invariance of the action} \label{GI:L_CS}
Here we check an invariance of the action 
under the gauge transformation~(\ref{dgt:superfield}). 
The invariance of the hypermultiplet sector is manifest. 
In the vector multiplet sector, the $d^4\tht$-integral part 
is also manifestly gauge-invariant since it consists of 
the field strength superfield~$\cV^I$ and the gauge singlet~$\tl{V}_E$. 
Hence the only nontrivial part is the (supersymmetric) Chern-Simons terms, 
\be
 \cL^{\rm CS} \equiv \int\dr^2\tht\;W^{\rm CS}+\hc, 
\ee
where $W^{\rm CS}$ is given by (\ref{expr:W^CS}). 

As explained in Sec.~\ref{vector_sector}, 
the dependences of $\cL^{\rm CS}$ 
on $U^\mu$ and $\Psi_\alp$ ($\bar{\Psi}_{\dalp}$) are not fixed 
completely only by the $\dscp$-invariance, \ie, 
a constant~$a$ in (\ref{expr:cZ^IJ}) remains undetermined. 
As we will show in the following, the gauge invariance determines its value. 

First, $\cL^{\rm CS}$ is split into three parts, \ie, 
$\cL^{\rm CS}_0\equiv\cL^{\rm CS}|_{U^\mu=\Psi_\alp=0}$, 
$U^\mu$-dependent part~$\Dlt_U\cL^{\rm CS}$, and 
$\Psi_\alp$($\bar{\Psi}_{\dalp}$)-dependent part~$\Dlt_\Psi\cL^{\rm CS}$. 
Their explicit forms are 
\bea
 \cL^{\rm CS}_0 \eql \int\dr^2\tht\;\frac{3C_{IJK}}{2}
 \brc{-\Sgm^I\cW_0^J\cW_0^K
 +\frac{\bar{D}^2}{12}\brkt{\cZ_0^{IJ\alp}}\cW_{0\alp}^K}+\hc, \nonumber\\
 \Dlt_U\cL^{\rm CS} \eql \int\dr^2\tht\;\frac{3C_{IJK}}{2}
 \left\{-\frac{1}{2}\Sgm^I\bar{D}^2\brkt{U^\mu\cW_0^J\sgm_\mu\bar{\cW}_0^K} 
 \right. \nonumber\\
 &&\left.\hspace{30mm}
 +\frac{\bar{D}^2}{12}\brkt{\Dlt_U\cZ^{IJ\alp}\cW_{0\alp}^K
 +\frac{1}{4}U^\mu\bar{D}^2\cZ_0^{IJ}\sgm_\mu\bar{\cW}_0^K}\right\}+\hc, 
 \nonumber\\
 \Dlt_\Psi\cL^{\rm CS} \eql \int\dr^2\tht\;\frac{C_{IJK}}{8}
 \brkt{\Dlt_\Psi\cZ^{IJ\alp}\cW_{0\alp}^K}+\hc, 
\eea
where 
\bea
 \cZ_0^{IJ\alp} \defa X^I D_\alp\der_y X^J
 -\der_y X^ID^\alp X^J, \nonumber\\
 \Dlt_U\cZ^{IJ\alp} \defa -\frac{1}{2}\bar{\sgm}^{\dbt\bt}_\mu U^\mu
 \left\{X^I D^\alp D_\bt\bar{D}_{\dbt}\der_y X^J
 +D_\bt\bar{D}_{\dbt}X^ID^\alp\der_y X^J \right. \nonumber\\
 &&\hspace{20mm}\left. 
 -\der_y X^I D^\alp D_\bt\bar{D}_{\dbt}X^J
 -D_\bt\bar{D}_{\dbt}\der_y X^ID^\alp X^J\right) \nonumber\\
 &&-\frac{1}{2}\bar{\sgm}^{\dbt\bt}_\mu D^\alp U^\mu
 \brkt{\bar{D}_{\dbt}X^ID_\bt\der_y X^J
 -\bar{D}_{\dbt}\der_y X^I D_\bt X^J} \nonumber\\
 &&-\frac{a}{4}\bar{\sgm}^{\dalp\alp}_\mu\der_y U^\mu
 X^I D^2\bar{D}_{\dalp}X^J, 
\eea
\bea
 \Dlt_\Psi\cZ^{IJ\alp} \defa -X^ID^\alp\brkt{
 \frac{1}{4}\bar{D}^2\Psi^\bt D_\bt+\frac{1}{2}\bar{D}^{\dbt}\Psi^\bt
 \bar{D}_{\dbt}D_\bt+\hc}X^J \nonumber\\
 &&+\brkt{\frac{1}{4}\bar{D}^2\Psi^\bt D_\bt
 +\frac{1}{2}\bar{D}^{\dbt}\Psi^\bt\bar{D}_{\dbt}D_\bt+\hc}X^ID^\alp X^J
 \nonumber\\
 &&-\frac{a}{4}\brkt{\bar{D}^{\dalp}\Psi^\alp
 -D^\alp\bar{\Psi}^{\dalp}}X^ID^2\bar{D}_{\dalp}X^J. 
\eea

The gauge transformation of $\cL_{\rm CS0}$ is calculated as 
\be
 \dgt\cL^{\rm CS}_0 = -\int\dr^2\tht\;\frac{3C_{IJK}}{2}
 \brc{\frac{1}{2}\der_y\brkt{\Lmd^I\cW_0^J\cW_0^K}
 -\cO_\Psi^{(1)}\Lmd^I\cW_0^J\cW_0^K}+\hc,  \label{dgt:cL0}
\ee
where 
\be
 \cO_\Psi^{(1)} \equiv \frac{1}{4}\bar{D}^2\Psi^\bt D_\bt
 +\frac{1}{2}\bar{D}^{\dbt}\Psi^\bt\bar{D}_{\dbt}D_\bt+\hc. 
\ee
We have used the partial integrals and 
$D^\alp\cW_{0\alp}^K=\bar{D}_{\dalp}\bar{\cW}_0^{K\dalp}$. 

After some calculations, 
the gauge transformation of $\Dlt_U\cL^{\rm CS}$ is obtained as 
\bea
 \dgt\Dlt_U\cL^{\rm CS} \eql -\int\dr^4\tht\;
 \frac{C_{IJK}}{2}\left[ U^\mu\der_y\brc{\brkt{\Lmd^I+\bar{\Lmd}^I}
 \bar{\cW}_0^J\bar{\sgm}_\mu\cW_0^K} \right. \nonumber\\
 &&\left.\hspace{20mm}
 +a\der_y U^\mu\brkt{\Lmd^I+\bar{\Lmd}^I}
 \bar{\cW}_0^J\bar{\sgm}_\mu\cW_0^K+\hc\right].  \label{dgt:DltUcL}
\eea
Here we have used the relation~$-\frac{1}{4}\bar{D}^2=d^2\bar{\tht}$ 
up to the total derivative, and the following identity that 
holds for arbitrary spinors~$\chi_\alp,\psi_\alp,\lmd_\alp$. 
\be
 \psi_\bt\chi^\bt\lmd^\alp-\psi^\alp\chi^\bt\lmd_\bt
 -\psi_\bt\chi^\alp\lmd^\bt = 0.  \label{rel:fermions}
\ee

Finally we consider the gauge transformation of $\Dlt_\Psi\cL^{\rm CS}$. 
\bea
 \dgt\Dlt_\Psi\cL^{\rm CS} \eql -\int\dr^4\tht\;\brc{
 \frac{C_{IJK}}{2}\dgt\Dlt_\Psi\cZ^{IJ\alp}\cW_{0\alp}^K+\hc} \nonumber\\
 \eql \int\dr^4\tht\;C_{IJK}\brc{
 \Lmd^I\brkt{\cO_\Psi^{(2)}}^\alp_{\;\;\bt}D^\bt X^J
 -\cO_\Psi^{(1)}\Lmd^I D^\alp X^J}\cW_{0\alp}^K+\hc,  \label{dgt:DltPsicL1}
\eea
where 
\bea
 \brkt{\cO_\Psi^{(2)}}^\alp_{\;\;\bt} \eql 
 \dlt^\alp_{\;\;\bt}\brkt{\frac{1}{4}\bar{D}^2\Psi^\gm D_\gm
 +i\sgm^\mu_{\gm\dgm}\bar{D}^{\dgm}\Psi^\gm\der_\mu} \nonumber\\
 &&-\brkt{\frac{1}{4}D^\alp\bar{D}^2\Psi_\bt
 +\frac{1}{2}D^\alp\bar{D}^{\dbt}\Psi_\bt\bar{D}_{\dbt}}. 
\eea
We have used (\ref{rel:DderXs}) in deriving (\ref{dgt:DltPsicL1}). 
By using (\ref{rel:fermions}), we can show that 
\be
 C_{IJK}\cO_\Psi^{(1)}\Lmd^I D^\alp X^J\cW_{0\alp}^K 
 = -2C_{IJK}\Lmd^I\brkt{\cO_\Psi^{(2)}}^\alp_{\;\;\bt}D^\bt X^J\cW_{0\alp}^K, 
\ee
up to total derivatives. 
Thus, we obtain 
\bea
 \dgt\Dlt_\Psi\cL^{\rm CS} \eql -\int\dr^4\tht\;
 \frac{3C_{IJK}}{2}\cO_\Psi^{(1)}\Lmd^I D^\alp X^J\cW_{0\alp}^K+\hc \nonumber\\
 \eql -\int\dr^2\tht\;\frac{3C_{IJK}}{2}\cO_\Psi^{(1)}\Lmd^I\cW^J_0\cW^K_0+\hc. 
 \label{dgt:DltPsicL2}
\eea

Summing up (\ref{dgt:cL0}), (\ref{dgt:DltUcL}) and (\ref{dgt:DltPsicL2}), 
we find that $\dgt\cL^{\rm CS}$ becomes the following total derivative 
when $a=1$. 
\bea
 \dgt\cL^{\rm CS} \eql -\int\dr^2\tht\frac{C_{IJK}}{2}
 \der_y\brc{\Lmd^I\cW_0^J\cW_0^K-\frac{1}{2}\bar{D}^2\brkt{
 U^\mu\Lmd^I\bar{\cW}_0^J\bar{\sgm}_\mu\cW_0^K}}+\hc \nonumber\\
 \eql -\int\dr^2\tht\;\frac{C_{IJK}}{2}
 \der_y\brkt{\Lmd^I\cW^J\cW^K}+\hc. 
\eea
Therefore, the gauge invariance of the action determines 
the value of $a$.

\subsection{$\bdm{\dscq}$-invariance of the action} \label{Action_invariance}
Here we show the invariance of the action 
under the $\dscq$-transformation in (\ref{dscq:sf}) 
at the zeroth order in the gravitational superfields. 
First we consider the hypermultiplet sector. 
From the definition of $\cU$ in (\ref{modify:cU}), 
$\cU(\Phi^{\oa})$ transforms as 
\bea
 \dscq\cU(\Phi_{\rm odd}) \eql \frac{1}{2}\brkt{Y+\bar{Y}}\der_y\Phi_{\rm odd}
 -\frac{i}{4}\bar{D}^2\brc{\tl{N}\brkt{e^{-V}}^t\bar{\Phi}_{\rm even}}
 +i\tl{N}\der_y\Phi_{\rm odd}, \nonumber\\
  \dscq\cU(\Phi_{\rm even}) \eql \frac{1}{2}\brkt{Y+\bar{Y}}\der_y\Phi_{\rm even}
 +\frac{i}{4}\bar{D}^2\brc{\tl{N}e^V\bar{\Phi}_{\rm odd}}
 +i\tl{N}\der_y\Phi_{\rm even}.
\eea
Then, $\Omg^{\rm hyper}$ and $W^{\rm hyper}$ in (\ref{cL^hyper:NA}) 
transform as  
\bea
 \dscq\Omg^{\rm hyper} \eql -\frac{1}{2}\der_y\sbk{
 \brkt{Y+\bar{Y}}\brc{\Phi_{\rm odd}^\dagger\tl{d}\brkt{e^V}^t\Phi_{\rm odd}
 +\Phi_{\rm even}^\dagger\tl{d}e^{-V}\Phi_{\rm even}}} \nonumber\\
 &&-\left[
 \frac{i}{4}\tl{N}\brkt{D^2\Phi_{\rm odd}^t\tl{d}\Phi_{\rm even}
 -\Phi_{\rm odd}^t\tl{d}D^2\Phi_{\rm even}}
 -\frac{i}{2}D^\alp\tl{N}\Phi_{\rm odd}^t\tl{d}D_\alp e^V 
 e^{-V}\Phi_{\rm even}\right.  \nonumber\\
 &&\hspace{10mm}
 +i\tl{N}\brc{\Phi_{\rm odd}^\dagger\tl{d}\brkt{e^V}^t\der_y\Phi_{\rm odd}
 +\Phi_{\rm even}^\dagger\tl{d}e^{-V}\der_y\Phi_{\rm even}} 
 \nonumber\\
 &&\left.\hspace{10mm}
 +i\tl{N}\brc{\Phi_{\rm odd}^\dagger\tl{d}\brkt{\Sgm e^V}^t\Phi_{\rm odd}
 -\Phi_{\rm even}^\dagger\tl{d}\brkt{e^{-V}\Sgm}\Phi_{\rm even}}+\hc \right], 
 \nonumber\\
 \dscq W^{\rm hyper} 
 \eql \der_y\sbk{Y\brc{\Phi_{\rm odd}^t\tl{d}\brkt{\der_y-\Sgm}\Phi_{\rm even}
 -\Phi_{\rm even}^t\tl{d}\brkt{\der_y+\Sgm^t}\Phi_{\rm odd}}} \nonumber\\
 &&-\frac{i}{4}\bar{D}^2\left[
 2\tl{N}\brc{\Phi_{\rm odd}^\dagger\tl{d}\brkt{e^V}^t\der_y\Phi_{\rm odd}
 +\Phi_{\rm even}\tl{d}e^{-V}\der_y\Phi_{\rm even}} \right. \nonumber\\
 &&\hspace{15mm}
 +2\tl{N}\brc{\Phi_{\rm odd}^\dagger\tl{d}\brkt{\Sgm e^V}^t\Phi_{\rm odd}
 -\Phi_{\rm even}^\dagger\tl{d}e^{-V}\Sgm\Phi_{\rm even}}  
 \label{dscq:Omg-W}  \\
 &&\hspace{15mm}\left. 
 -\frac{1}{2}\brkt{\Phi_{\rm odd}^t\tl{d}D^2\Phi_{\rm even}
 -\Phi_{\rm even}^t\tl{d}D^2\Phi_{\rm odd}}
 -D^\alp\tl{N}\Phi_{\rm odd}^t\tl{d}D_\alp e^V e^{-V}\Phi_{\rm even}\right], 
 \nonumber
\eea
where we have dropped total derivatives for $x^\mu$ 
and $\tht_\alp$ ($\bar{\tht}_{\dalp}$). 
Thus, we can see that 
\be
 \dscq\sbk{2\int\dr^4\tht\;\Omg^{\rm hyper}
 +\brc{\int\dr^2\tht\;W^{\rm hyper}+\hc}} = 0, \label{hyper:cancel}
\ee
up to total derivatives, by using $-\frac{1}{4}\bar{D}^2=d^2\bar{\tht}$. 
Note that the cancellation occurs between the $\dscq$-variations of 
the $d^4\tht$-integral and the $d^2\tht$-integral. 

Next we consider the vector multiplet sector. 
For simplicity, we assume that the gauge group is Abelian. 
From (\ref{dscq:sf}) and (\ref{dscq:cWV}), we see that 
\be
 \dscq\brkt{\frac{\cV^I}{V_E}} = \frac{1}{2}\brkt{Y+\bar{Y}}\der_y\cV^I
 -\brc{\frac{i}{2}D^\alp\brkt{N\cW_\alp^I}+\hc}.  \label{dscq:cV/V_E}
\ee
The ratio~$\cV^I/V_E$ is identified with 
an $\cN=1$ real general multiplet~(4.19) in Ref.~\cite{Kugo:2002js}. 
The first term in (\ref{dscq:cV/V_E}) corresponds to 
(4.15) of Ref.\cite{Kugo:2002js}. 
Then we obtain the transformation law of $\Omg^{\rm vector}$ 
defined in (\ref{expr:Omg^vector}) as 
\be
 \dscq\Omg^{\rm vector} = \der_y\brc{\frac{1}{2}\brkt{Y+\bar{Y}}
 \Omg^{\rm vector}}-\brc{\frac{3}{8}C_{IJK}\cV^I\cV^J
 D^\alp Y\cW_\alp^K+\hc}
\ee
Since it is cumbersome to show the invariance of the whole action 
in this sector, here we focus on the quadratic terms in $\Sgm^I$ 
in the $\dscq$-variation to illustrate the cancellations. 
Then, $W^{\rm CS}$ defined in (\ref{expr:W^CS}) transforms as 
\bea
 \dscq\brc{\int\dr^2\tht\;W^{\rm CS}+\hc} 
 \eql \int\dr^2\tht\;\frac{3C_{IJK}}{2}\brkt{
 -2\Sgm^I\dscq\cW^J\cW^K+\cdots}+\hc \nonumber\\
 \eql -\int\dr^2\tht\;\frac{3C_{IJK}}{8}\bar{D}^2\sbk{
 \Sgm^I D^\alp\brc{\brkt{Y+\bar{Y}}\cV^J}\cW^K_\alp+\cdots}+\hc \nonumber\\
 \eql \int\dr^4\tht\;\frac{3C_{IJK}}{4}\sbk{
 D^\alp Y\brkt{\Sgm^I+\bar{\Sgm}^I}\brkt{\Sgm^J+\bar{\Sgm}^J}\cW_\alp^K
 +\cdots+\hc}, \nonumber\\
\eea
where the ellipses denote terms up to linear in 
$\Sgm^I$ or $\bar{\Sgm}^I$.  
We have dropped total derivatives, and used $-\frac{1}{4}\bar{D}^2=d^2\bar{\tht}$
and the fact that $D^\alp\cW_\alp^I$ is real at the zeroth order 
in the gravitational superfields.  
Therefore, we can see the cancellation between the $\dscq$-variation of 
the $d^4\tht$-integral and the $d^2\tht$-integral, 
\be
 \dscq\sbk{2\int\dr^4\tht\;\Omg^{\rm vector}
 +\brc{\int\dr^2\tht\;W^{\rm CS}+\hc}} = 0,  \label{vector:cancel}
\ee
up to total derivatives. 

Summing up (\ref{hyper:cancel}) and (\ref{vector:cancel}), 
we obtain (\ref{cancellation}).


\end{document}